\documentclass{article}

\usepackage[main,final]{neurips_2026}

\usepackage[utf8]{inputenc} 
\usepackage[T1]{fontenc}    
\usepackage{hyperref}       
\usepackage{url}            
\usepackage{booktabs}       
\usepackage{amsfonts}       
\usepackage{nicefrac}       
\usepackage{microtype}      
\usepackage{xcolor}         
\usepackage{amsmath}
\usepackage{graphicx}
\usepackage{multicol,multirow}
\usepackage{wrapfig}
\usepackage{amsmath, inputenc}
\usepackage{tcolorbox}
\usepackage{algorithm}
\usepackage{algorithmic}
\usepackage{tikz}
\usepackage{pifont}
\newcommand{\cmark}{\ding{51}}
\newcommand{\xmark}{\ding{55}}

\definecolor{evblue}{HTML}{3b82f6}
\definecolor{evred}{HTML}{ef4444}
\definecolor{evgreen}{HTML}{22c55e}
\definecolor{evamber}{HTML}{f59e0b}
\definecolor{evviolet}{HTML}{a855f7}
\definecolor{evcyan}{HTML}{06b6d4}
\definecolor{lightgray}{gray}{0.7}
\newcommand{\gc}{\textcolor{gray}{\cmark}}  
\newcommand{\g}[1]{\textcolor{lightgray}{#1}}    

\usepackage[textsize=tiny]{todonotes}
\newcommand{\method}{\textsc{TAC}}

\usepackage{hyperref}

\usepackage{subcaption}

\title{TAC: Timestamped Audio Captioning}

\author{%
Sonal Kumar$^{1,2,*}$ \And Prem Seetharaman$^{2,*}$ \And Ke Chen$^2$ \And Oriol Nieto$^2$ \And Jiaqi Su$^2$ \And
Zhepei Wang$^2$ \And Rithesh Kumar$^{3,\dagger}$ \And Dinesh Manocha$^1$ \And Nicholas J. Bryan$^2$ \And
Zeyu Jin$^2$ \And Justin Salamon$^2$ \\
\small $^1$University of Maryland, College Park, USA \quad $^2$Adobe Research, USA \\
\small $^3$OpenAI, USA \quad $^*$Equal contribution \quad $^\dagger$Work done while at Adobe \\
\small \texttt{sonalkum@umd.edu, seethara@adobe.com}
}

\begin{document}

\maketitle

\begin{abstract}
Large Audio Language Models struggle to disentangle overlapping events in complex acoustic scenes, yielding temporally inconsistent captions and frequent hallucinations. We introduce Timestamped Audio Captioner (\method), a model that produces temporally grounded audio descriptions at varying degrees of detail and resolution. \method{} is trained with a synthetic data pipeline that constructs challenging and dynamic mixtures from real-world audio sources, enabling robust learning under realistic polyphonic conditions. Across event detection, dense captioning, and speaker diarization, \method{} outperforms most competing methods - including dedicated diarization systems with a low hallucination rate and accurate temporal grounding. We also introduce \method{}-V, an audio-visual pipeline to generate semantically rich audio-visual descriptions. We then show that \method{} and \method{}-V serves as a ``semantic bridge'' for a text-only reasoner: a simple \method$\rightarrow$LLM and \method{}-V$\rightarrow$LLM cascade achieves state-of-the-art scores on benchmarks for both audio (MMAU-Pro, MMSU, MMAR) and audio-visual (DailyOmni, VideoHolmes) understanding and reasoning respectively. We encourage readers to see detailed qualitative results on our demo page: \url{https://sonalkum.github.io/tacmodel/}.
\end{abstract}

\section{Introduction}
\label{sec:intro}
\begin{wrapfigure}{r}{0.5\textwidth}
\vspace{-2em}
\centering
\includegraphics[width=\linewidth, trim=0 0 0 0.5cm]{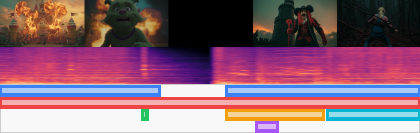}
\small
\raggedright
Audio $\rightarrow$ \method{} $\rightarrow$ \textbf{{\color{evblue}[music] Heroic brass fanfares and thunderous percussion from 0.0s to 3.8s, 5.4s to 10.0s.} {\color{evred}[sfx] Fire crackling and burning from 0.0s to 10.0s.} {\color{evgreen} Sudden burst of sound from 3.4s to 3.5s.} {\color{evamber}[sfx] A group of people shouting in unison, expressing excitement from 5.4s to 7.7s.}  {\color{evviolet}[sfx] Heavy object crashes down from 6.1s to 6.6s.} {\color{evcyan}[sfx] Rattling and clattering from a moving chain from 7.8s to 10.0s.}}
\caption{\small Given only audio, \method{} generates structured, timestamped descriptions of overlapping sound events. We visualize the timestamps produced by \method{} as temporal lanes above. Colors indicate correspondence between text and temporal lanes.}
\label{fig:tac-example-output}
\vspace{-1em}
\end{wrapfigure}
The pursuit of \emph{audio general intelligence} is rapidly advancing with Large Audio-Language Models (LALMs), which promise to turn raw audio into rich semantic understanding for captioning, instruction following, and open-ended reasoning. Recent foundation models including SALMONN~\citep{tang2024salmonn}, Qwen2-Audio~\citep{chu2024qwen2audio}, GAMA~\citep{gama}, the Audio Flamingo series~\citep{kong2024audio,af2,af3}, Audio-Thinker~\cite{thinker}, Kimi-Audio~\cite{kimi}, and MiMo-Audio~\cite{mimo} have demonstrated impressive progress across speech, sound, and music understanding. Yet, when deployed on complex real-world auditory scenes with \emph{overlapping} and \emph{time-varying} events, these systems remain brittle. Even strong frontier models (e.g., Gemini 3 Pro~\citep{gemini25}) often produce \emph{global} captions that miss fine-grained temporal structure, confuse event boundaries, or hallucinate non-existent sounds -- failure modes that recent benchmarks and analyses identify as central obstacles to reliable audio understanding~\citep{kuan2024can,cheng2025ahabench}.

We argue that these failures reflect a fundamental \emph{supervision mismatch} between continuous, high-density audio streams and the sparse language annotations used to train LALMs.
Popular captioning datasets (e.g., AudioCaps~\citep{kim2019audiocaps}, Clotho~\citep{drossos2020clotho}) typically provide a single caption for a 10--30 second clip.
This results in \emph{semantic collapse}: temporally distinct events are compressed into a short, clip-level summary, making it difficult for models to preserve causality and disentangle overlaps. Language priors can then dominate and yield hallucinations~\citep{kuan2024can,cheng2025ahabench}. Recent alignment efforts further suggest that grounding failures are systemic, and can be reduced only when training includes hard counterfactual negatives targeting fine-grained temporal reasoning~\citep{cheng2025ahabench}. These findings indicate that robust audio understanding requires bridging dense audio with \emph{structured, temporally grounded} linguistic supervision.

We propose Timestamped Audio Captioner (\method), a model trained to produce timestamped audio description (see Fig. \ref{fig:tac-example-output}). \method{} produces captions paired with exact start and end times for every source in complex auditory scenes. Unlike prior LALMs which tackle broader understanding and reasoning \cite{af2, af3, gama, gemini25, Qwen3Omni}, \method{} focuses on ``what happens when'' (e.g. sound event detection). We then cascade \method{} with a ``reasoner'' (a text-only LLM), resulting in a ``describe-then-reason'' approach to multimodal understanding. From audio, \method{} produces high-quality dense text captions that serve as evidence that the reasoner uses to answer questions. Finally, we extend this to audiovisual inputs by pairing \method{} with an off-the-shelf VLM. Remarkably, we find that this simple cascade obtains state-of-the-art results on several multimodal understanding benchmarks. By decoupling the describer from the reasoner, we can scale the two components independently. We show that stronger reasoners give higher performance, even when given access to the same \method{} descriptions. 

While our architecture follows the standard LALM paradigm, our key insight is that the bottleneck lies in \emph{supervision granularity}, not model complexity. We show that a principled data pipeline with dynamic supervision including a weighted loss (Eq.~\ref{eq:weight_time_loss}) and multitask prompt conditioning enables a standard backbone to surpass far larger models trained on orders of magnitude more data. Our contributions are: (i) \textbf{\method{}:} an audio understanding model trained on a synthetic, multi‑granular curriculum generated by a dynamic data generation pipeline, achieving SOTA results in audio captioning, sound event detection (SED) and speaker diarization tasks; (ii) \textbf{\method{}-V:} an audio‑visual extension obtained by pairing \method{} with a vision–language model to produce dense audio‑visual captions; and (iii) \textbf{Describe then reason:} dense captions from \method{}(-V) are a semantic bridge for reasoning with text‑only LLMs, yielding state-of-the-art performance on audio reasoning benchmarks (\textit{MMAR}~\cite{mmar}, \textit{MMSU}~\cite{wang2025mmsu}, \textit{MMAU-Pro}~\cite{kumar2025mmau}) and competitive results on \textit{MMAU}~\cite{sakshi2024mmau}, as well as state-of-the-art or competitive audiovisual reasoning performance when combining \method{}-V with a text-only LLM (\textit{DailyOmni}~\cite{dailyomni}, \textit{VideoHolmes}~\cite{holmes}, \textit{WorldSense}~\cite{worldsense}, \textit{AVHBench}~\cite{avhbench}).

\section{Related Work}
\textbf{LALMs.}
Recent work in audio perception and understanding has shifted from task-specific models \citep{gong2021ast, chen2022beats} to general-purpose generative systems. Works like LTU \citep{sgong2023listen} and SALMONN \citep{tang2024salmonn} demonstrated that aligning audio encoders (e.g., Whisper, AudioMAE) with LLMs enables zero-shot speech and audio reasoning. Instruction-tuned models, such as GAMA~\cite{gama}, Qwen-Audio \citep{chu2023qwenaudio} and Audio Flamingo series\citep{kong2024audio, af2, af3}, have scaled  approach, achieving impressive performance by embedding audio directly into the context of an LLM. AudioChat~\cite{audiochat} enables audio foundation models to generate, edit, and understand complex ``audio stories" (multi-speaker, multi-source scenes) by simulating realistic training data with LLM agents and training with Audio Transfusion Forcing. However, these models often falter in ``cocktail party'' scenarios involving overlapping sound events. Even strong frontier models like Gemini 3 Pro \citep{gemini25} still remain prone to hallucinating events not present in the audio \citep{kuan2024can}. We attribute this to the ``global pooling'' nature of their supervision, where temporal details are compressed into a single semantic vector. In contrast, \method{} enforces a dense, time-aware alignment, enabling detailed reasoning.

\textbf{Audio and Audiovisual Captioning.}
Automated Audio Captioning (AAC) has traditionally relied on human-annotated datasets like AudioCaps \citep{kim2019audiocaps} and Clotho \citep{drossos2020clotho}, limited by fewer samples and ``sparse'' annotation—a single sentence for a 10--30\,s clip. This forces models to learn correlations rather than causality. While dense captioning is well-explored in vision \citep{johnson2016densecap}, it remains nascent in audio due to the cost of dense timestamp annotation. Weakly-supervised approaches like WavCaps \citep{mei2024wavcaps} scale up using noisy metadata but lack precise temporal boundaries for SED. Although AudioSet-Strong~\cite{audiosetstrong} offers timestamped event labels and TACOS~\cite{primus2025tacos} targets temporal alignment with human-annotated clips, their focus is atomic classification and free-text SED, not dense descriptions. In the audiovisual domain, Video-SALMONN~2~\cite{videosalmonn2} extends SALMONN with caption-enhanced training, while Omni-Captioner~\cite{omnicaptioner} and AuroraCap~\cite{auroracap} target omni-modal and video-level dense perception. \method{} addresses annotation scarcity by synthesizing a curriculum of dense, temporally-precise captions bridging simple tagging and complex storytelling.

\textbf{Synthetic and LLM-based Data Generation for Audio.}
Recent works use LLMs to generate QA pairs or captions from audio metadata. GAMA \citep{gama} and Audio Flamingo 2/3 \citep{af2, af3} utilize GPT-4 for complex QA and reasoning chains, while ReCLAP \citep{ghosh2025reclap} augments training data by rewriting captions to emphasize acoustic characteristics. These focus on global clip-level understanding but lack fine-grained temporal grounding. Scaper~\citep{salamon2017scaper} mixes isolated events for closed-vocabulary SED; FLAM~\citep{wu2025flam} mixes audio to train a contrastive frame-wise discriminator; and Hive~\citep{hive} synthesizes semantically consistent mixtures for query-based universal sound separation. Unlike these objectives, \method{} generates multi-granular, open-vocabulary caption supervision with exact timestamps and jointly sampled style, resolution, merge, and activity controls. In the audiovisual domain, AVoCaDo~\cite{avocado} generates temporally orchestrated captions using Gemini~2.5-Pro with GPT-4.1 filtering, while Omni-Detective~\cite{omnicaptioner} orchestrates multiple frontier models with up to 10 iterative rounds per sample. \method{}-V instead uses a single TAC pass followed by a single VLM pass.

\section{Methodology}
\label{sec:method}

\begin{wrapfigure}{r}{0.5\textwidth}
    \centering
    \vspace{-4em}
    \includegraphics[width=0.5\textwidth, trim= 0 0.3cm 0 0cm]{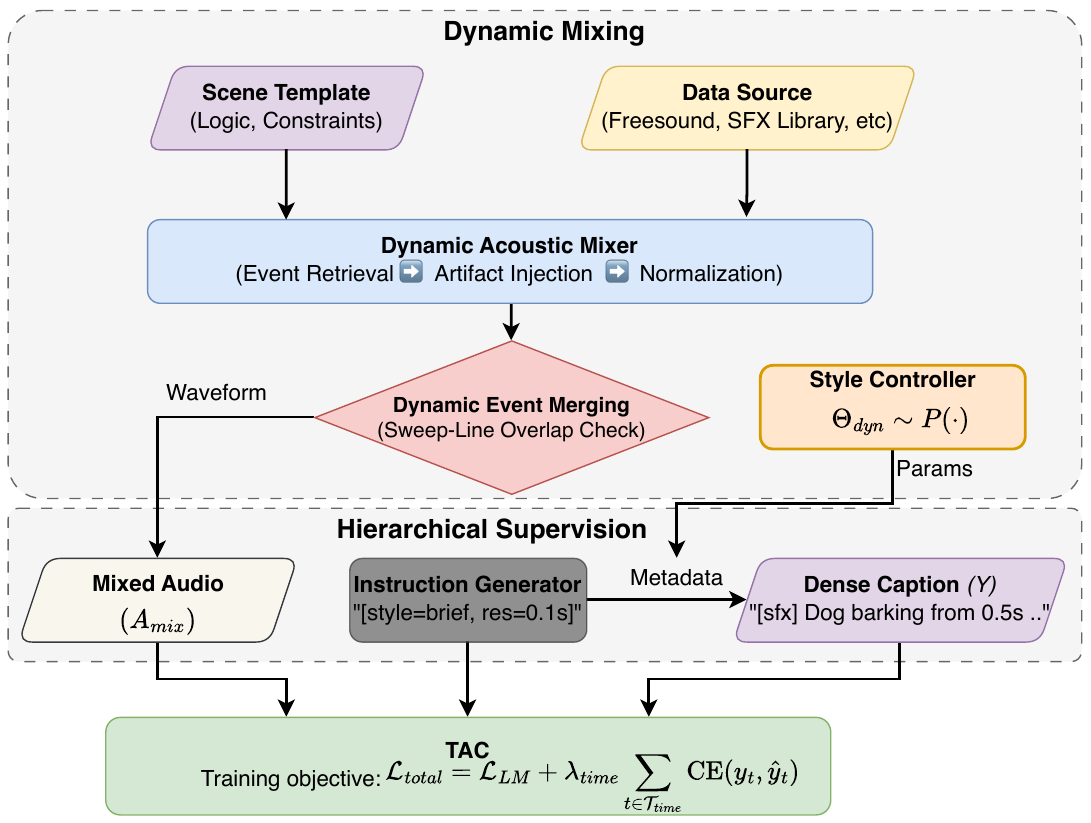}
    \caption{\small\textbf{The \method{} Training Pipeline.} Stage 1 synthesizes complex audio mixtures via our \textit{Dynamic Acoustic Mixer}. In Stage 2, a \textit{Style Controller} stochastically samples ``description styles" (Keyword vs. Brief vs. Detailed) and timing resolutions, generating a diverse curriculum of instruction-tuned prompts.
    }
    \label{fig:tac-train-pipeline}
    \vspace{-2em}
\end{wrapfigure}

We introduce \method, a model designed to bridge the gap between low-level acoustic signals and high-level reasoning.
This pipeline allows us to finetune a standard LALM~\citep{chu2023qwenaudio} to achieve state-of-the-art dense captioning within just 5k training iterations over synthetic mixtures.
The proposed methodology is depicted in Figure~\ref{fig:tac-train-pipeline}, and below we detail all its respective steps.

\subsection{Dynamic Acoustic Mixer}
While recent works scale model size to improve performance, we argue that the bottleneck lies in the \textit{granularity} of supervision. Standard datasets provide a single ``global" caption for a complex scene, forcing models to average out temporal details. 
To overcome this, we use a \textbf{Dynamic Acoustic Mixer} that generates infinite, highly-complex audio mixtures with synchronized ground truth at multiple levels of semantic resolution from single-source audio datasets.To increase the realism of the mixer, we define acoustic scenes via \textbf{Scene Templates} that specify the structural logic of an audio clip. A template $T$ consists of a set of temporal constraints $C$ and role bindings $R = \{r_{speech}, r_{music}, r_{sfx}, r_{bg}\}$. For example, a ``Speech over Music in Indoor Environment" template might require that the music source plays continuously, a speech source can occur randomly throughout (while never overlapping with another speech stream), and the sound effects source is restricted to background ambience, keyboard clicking, phone ringing. While the actual underlying sources are random, by tuning these templates we can make an endless combination of targeted synthetic mixtures for specific tasks. Our mixer allows for flexible control of various properties, such as number of concurrent sounding events, amount of reverberation and other signal-level augmentation, and number of repeats of an event.

Finally, precise temporal grounding is achieved via RMS-based activity detection with an activity threshold of $\delta_{act}$ (a proxy for loudness) unlike metadata which is often used in literature and relevant works. For every instantiated event $e_i$, we compute a continuous activity map $M_i(t)$. Given a merge threshold $\delta_{\text{merge}} \sim \mathcal{U}(0.1,\, 1.0)$, in seconds, if two activations of the same event are separated by a gap $g < \delta_{\text{merge}}$, they are fused into a single timestamped segment. While one can choose $\delta_{act}$ and $\delta_{\text{merge}}$ statically before training, we instead choose them per example during training, and condition the model on the chosen values.

In practice, we generate ${\sim}$150K synthetic mixtures (${\sim}$1.2K hours) for training, with an average of 5 events per clip and ${\sim}$20s average clip duration. The full mixing procedure is described in Algorithm~\ref{alg:scene_compilation}, and a comparison with existing temporally grounded audio-caption resources is provided in Table~\ref{tab:dataset_comparison} (both in Appendix~\ref{sec:dataset_details}).
\vspace{-1em}

\subsection{Multitask prompts and output format}
Instead of fixing the tasks statically at the beginning of the training (for example deciding that model must detect sounds with a granularity of $0.25$s), we instead sample from a set of multitask prompts, and modify the target caption accordingly. There are 4 high-level properties for each task that we can control per training sample:

\noindent\textbf{1. Style}: we sample from various caption styles for each event in the soundscape. These styles can be brief (``Dog barks''), keywords (``Dog''), or detailed (``A dog barks aggressively twice'').

\noindent\textbf{2. Merge threshold}: $\delta_{\text{merge}}$ dictates how close an events offset must be near its closest onset before they are merged into one item. For example, this can decide if two quick utterances are detected as one event (e.g. ``Speech from $5.0$s to $10.0$s'', or two events (e.g. ``Speech from $5.0$s to $7.0$s, $8.0$s to $10.0$s'').
\begin{wrapfigure}{r}{0.5\textwidth}
\vspace{-1.5em}
\small
\centering
\begin{tcolorbox}[colback=gray!10!white, colframe=black!75!black, title=\textbf{Exemplar Training Pair}]
\small
\textbf{Input Audio (Synthetic Mixture):} \\
\textit{[0.0-10.0s] Background Jazz Music + [0.5-2.3s] Dog Bark + [5.0-5.5s] Car Horn}

\vspace{0.1cm}
\textbf{Conditioning Prompt:} \\
\texttt{<|im\_start|>user Describe all events in the audio. Give start and end times. [style=brief, merge=0.25s, activity=0.05, resolution=0.10s] <|im\_end|>}

\vspace{0.1cm}
\textbf{Target Output (Ground Truth):} \\
\texttt{<|im\_start|>assistant 3 events total. 2 events overlap. 2 sound effects, 1 music. [music] Background jazz music plays softly from <|0.00|>s to <|10.00|>s. [sfx] A dog barks aggressively from <|0.50|>s to <|2.30|>s. [sfx] A car horn honks briefly from <|5.00|>s to <|5.50|>s.<|im\_end|>}
\end{tcolorbox}
\caption{\small An example of a synthetically generated training pair. Note how the ``Reasoning Header" (``3 events total...") is algorithmically derived from the composition metadata, teaching the model to summarize before detailing.}
\label{fig:data_example}
\vspace{-3em}
\end{wrapfigure}
\noindent\textbf{3. Activity threshold}: $\delta_{act}$ controls how quiet a sound must get to its minimum before it is considered ``off''. This has an effect on sounds that are intermittent, but do not go all the way to silence, such as explosions, whooshes, or other sound design elements. A high activity threshold will break up sounds into many events; a low activity threshold will keep them as one.

\noindent\textbf{4. Time resolution}: We round off start and end times randomly when deciding what ground truth is. For example, we can round off to the nearest half second, or tenth of a second. This controls the resolution at which we want to caption the audio.

As shown in Algorithm \ref{alg:scene_compilation}, during training, we randomly sample a \textit{Caption Style} $\mathcal{S} \in \{\textsc{Keywords}, \textsc{Brief}, \textsc{Detailed}\}$ and a set of \textit{Timing Parameters} (resolution $\delta_{res}$, merge threshold $\delta_{merge}$, and activity threshold $\delta_{act}$).
The instruction prompt $P$ is conditioned on these parameters (e.g., \texttt{"[style=brief, resolution=0.1s]"}). This instruction tuning allows us to control the model's output density at inference time. We form the target sequence in a token efficient way by concatenating all start and end times for each event as a comma separated list with the description. Captions are ordered by start time. Each caption is associated with a ``type'' (music, sfx, speech, background), which is prepended to the caption as `[type]`. An example of an input/output pair can be seen in Figure~\ref{fig:data_example}. The structured output of \method{} can be easily parsed into a data structure, and used reliably for downstream tasks.

\subsection{\method{} Architecture and Training}
Though any backbone can be used, we use Qwen2-Audio~\citep{chu2023qwenaudio} for ours, freezing the base model and fine-tuning via Low-Rank Adaptation (LoRA)~\citep{hu2022lora} on linear layers. Standard LALMs, including our backbone Qwen2-Audio, are trained on broad in-the-wild data. While effective for general audio, they miss fine-grained, domain-specific acoustics (e.g., distinguishing an ``industrial hum'' from a ``sci-fi drone''), undermining dense captioning. Therefore, we decided to continue pretraining on Qwen2Audio on a large corpus of high-fidelity licensed single-source audio (e.g. an explosion sound effect, or a music track) paired with descriptive captions at varying levels of detail. We generated these captions from metadata, following the approach laid out in AudioCards~\cite{sridhar2026audiocards}. We expanded these captions to an instruction tuning set using off-the-shelf LLMs ({GPT-OSS-120B}~\cite{gptoss120b} and {Qwen-32B-VL}~\cite{qwen3vl}) with a variety of questions, such as identification (``What is the source of this sound?''), and description (``Describe the mood.'').

Standard cross-entropy loss is often insufficient for dense captioning, as it treats short-duration timestamp tokens equally with semantic tokens. To strictly enforce temporal precision, we tokenize timestamps as atomic special tokens (e.g., \texttt{<|1.23|>}), as done in prior work \cite{radford2023whisper, chu2023qwenaudio}. Unlike prior work, we introduce a weighted loss objective $\mathcal{L}_{total}$:
\begin{equation}
    \mathcal{L}_{total} = \mathcal{L}_{LM} + \lambda_{time} \sum_{t \in \mathcal{T}_{time}} \text{CE}(y_t, \hat{y}_t) + \lambda_{spk} \sum_{s \in \mathcal{T}_{spk}} \text{CE}(y_s, \hat{y}_s)
    \label{eq:weight_time_loss}
\end{equation}
where $\mathcal{T}_{time}$ represents the set of indices corresponding to timestamp tokens, $\mathcal{T}_{spk}$ the indices of speaker identity tokens (e.g., ``Speaker 1'', ``Speaker 2''), $\lambda_{time}$ upweights temporal alignment errors, and $\lambda_{spk}$ upweights speaker identity token errors during speaker diarization fine-tuning ($\lambda_{spk}{=}0$ for captioning/SED, $\lambda_{spk}{=}10.0$ for diarization; see Section~\ref{subsec:diarization}). Finally, while \method{} can be directly trained for speech transcription, we transcribe detected speech separately. Each TAC-detected speech interval is cropped and passed independently to Whisper~\cite{radford2023whisper}; TAC supplies timestamps and speaker attribution, while Whisper supplies only text, so no cross-model timestamp alignment is required. For overlap, TAC emits separate events and crops per speaker, supervised by the Overlapping Speakers template. If voices cannot be separated acoustically, transcription degrades as in other ASR systems and we do not claim reliable per-speaker transcripts, although event boundaries and attribution still come from TAC. For example, ``Male voice whispering from $1.0$s to $8.0$s'' expands to ``Male voice whispering from $1.0$s to $8.0$s {\small\ttfamily<speech>}Do you want to know a secret?{\small\ttfamily</speech>}''.

\subsection{\method{}-V: \method{} with Visuals}
\label{sec:av_pipeline}
\begin{wrapfigure}{r}{0.5\textwidth}
\vspace{-4em}
\centering
\begin{tcolorbox}[colback=gray!10!white, colframe=black!75!black, title=\textbf{TAC-V: Fused Audio-Visual Captions}]
\small
\fontfamily{pcr}\selectfont 
1. [0.2s - 2.0s] [music] Epic orchestral surge as metallic studio logo glows (0.99)\\
2. [3.8s - 6.2s] [sfx] Explosion erupts as debris flies through smoke (0.83)\\
3. [7.4s - 9.0s] [speech] Woman (whispering, on-screen) with furrowed brow (0.89)\\
\hspace*{2em}<speech>"Their cruelty is all I've known."</speech>
\end{tcolorbox}
\caption{An example output from our cascaded Audio-Visual pipeline. Note the integration of visual details (``metallic studio logo'', ``furrowed brow'') with precise audio events, and the inclusion of FLAM confidence scores (e.g., $0.99$) alongside aligned transcriptions.}
\label{fig:av_example}
\vspace{-2em}
\end{wrapfigure}
To demonstrate the extensibility of \method{}, we introduce \textbf{\method{}-V}, a pipeline that fuses the high temporal-precision outputs of \method{} with a Visual Language Model (VLM) for temporally dense audio-visual captions. The pipeline processes audiovisual inputs to produce timestamped, visually-grounded captions via five distinct stages. We first extract the audio and sample video frames at a configurable frame rate (we choose $2$fps). For video resolution, we alternate between 360p and 240p for every other frame, to stay within the effective token limit of our chosen VLM. 

\textbf{Audio captioning:} We process the audio by chunking it into $20s$ non-overlapping chunks. Each chunk is processed in parallel with \method{}. Unlike other audio LMs, \method{} provides precise timestamped events tagged by category (e.g., \texttt{[speech]}). We augment the output of \method{} by transcribing all detected speech events. Finally, we score every event by using {FLAM}~\citep{wu2025flam}, which assigns a confidence score $c \in [0, 1]$ to each detected event. This serves as a signal for the downstream VLM: low confidence scores flag ambiguous events that require visual verification.

\textbf{Audio-driven video captioning:} From \method{}, we create a ``shot-list'' of audio events, ordered by time, with precise timestamps, types, captions, and transcriptions. We augment this shot-list with visual shot boundaries (points where the scene changes in the video), placing them in the scene as visual markers. This ensures even coverage across an entire video, and aids the model in distinguishing continuous audio events from changing visual perspectives. We feed the video frames, the timestamped shot-list and confidence scores into {Qwen3-VL-32B}. Using a specialized \textit{Chain-of-Thought} prompt, the VLM performs \textit{Hallucination Correction} (using visuals to resolve acoustic ambiguity) and \textit{Visual Grounding} (linking sounds to visible sources). Figure~\ref{fig:av_example} illustrates the final structured output; the complete prompt and worked example appear in Appendix~\ref{sec:vlm_system_prompt}. The fused captions combine acoustic classification (e.g., \texttt{[sfx]}), visual grounding (e.g., ``debris flies''), and speech transcription into a unified timeline.
\subsection{Evaluation}
\label{subsec:evaluation}

Evaluating dense audio captioning is challenging because a single acoustic scene can be validly described at multiple levels of granularity, making standard metrics such as CIDEr~\cite{cider}, SPICE~\cite{spice}, and SPIDEr~\cite{liu2017improved} insufficient for capturing temporal precision or factual correctness. To address this limitation, we evaluate \method{} along three axes: \textit{semantic alignment}, \textit{temporal precision}, and \textit{robustness}.

\textbf{Semantic alignment:} Exact string matching is insufficient for dense captions~\cite{kumar2025mmau} (e.g., ``car engine" vs. ``vehicle idling" should be a match). We propose a reference-based metric using an LLM as a judge. For every predicted event $e_{pred}$ and ground truth event $e_{gt}$, we compute a \textit{Semantic Similarity Score} $S_{sem} \in [0, 1]$:
\begin{equation}
    S_{sem}(e_{pred}, e_{gt}) = \text{LLM}_{\text{judge}}(d_{pred}, d_{gt})
    \label{eq:semali}
\end{equation}
We use GPT-OSS-120B~\cite{gptoss120b} at temperature zero as the judge. It scores each description pair from 0 to 10 using a four-band rubric with worked examples, and we normalize the result to $[0,1]$. Matching is one-to-one and greedy: a candidate pair must have semantic similarity of at least $0.5$ and either temporal overlap or containment; eligible candidates are ordered by semantic similarity weighted by temporal overlap. Thus semantic agreement alone cannot produce a match. We additionally report sentence-embedding-based FENSE, alongside METEOR, CIDEr, and SPICE, to complement the compositional LLM judgment.
\textbf{Temporal precision:} To rigorously test the model's ability to localize events, we adapt Sound Event Detection (SED) metrics~\citep{app6060162,temko2006clear}.
After semantic alignment with a ground truth reference caption, we treat the generated captions as detection outputs and compute:

\noindent\textbf{(i) Segment-Based F1 (SegF1):} Evaluates activity detection at a $100$ms resolution. This measures how well the predicted duration matches the ground truth, regardless of the exact start/end times.

\noindent\textbf{(ii) Event-Based F1 (EvtF1):} Treats each caption segment as a discrete event. A prediction is counted as a True Positive (TP) only if its onset is within a $\pm 1.0$s window (or \emph{collar}) of the ground truth onset.

\textbf{Robustness \& Hallucination:} Hallucination remains a major challenge for LALMs~\cite{chen2025aha}. These models frequently produce temporally misaligned descriptions, invent subtle sound effects, misinterpret overlapping events, or confuse acoustically similar sources.
To assess performance in the absence of ground truth (or to detect hallucinations where the ground truth is silent), we utilize FLAM~\citep{wu2025flam} for reference-free evaluation. Reference matching remains primary for EvtF1, SegF1, and semantic similarity, but incomplete annotations can mark an unannotated-yet-present sound as a false positive and cannot distinguish it from an acoustically absent invention; FLAM complements it by scoring each predicted event directly against the audio. We define the \textit{Hallucination Rate (Hal\%)} as the percentage of predicted events where the FLAM confidence score drops below $\tau=0.25$, fixed before evaluation. A sweep from $0.15$ to $0.50$ confirms that the ranking is stable from $\tau=0.20$ upward (Appendix~\ref{sec:rebuttal_results}). We report \textbf{confidence (conf)} - the maximum audio-text similarity within the predicted time range - and \textbf{specificity (spec)} - the \textit{minimum} similarity across the predicted range. A high specificity indicates the model is not just detecting a peak, but accurately describing the entire duration of the event. On 100 human-rated TACOS samples, the LLM semantic judge correlates with human ratings at Spearman $0.81$/Kendall $0.76$, while FLAM confidence reaches $0.73/0.69$ (Appendix~\ref{sec:rebuttal_results}).

\section{Experiments}
\label{sec:experiments}
\noindent\textbf{Training Setup.} We train \method{} on a cluster of 8 NVIDIA A100 (80GB) GPUs, with a global effective batch size of 32. We freeze the pre-trained backbone and only fine-tune low-rank adapters (LoRA) with a rank of r=128 and alpha $\alpha$=256. Optimization is performed using AdamW with a peak learning rate of 5e-5, following a cosine decay schedule with $1000$ steps of linear warmup. The training data is a multi-stream mixture comprising our synthetic audio mixtures, SED data, TACOS, and real-world diarization data (${\sim}$943 hours from 10 corpora across 6 languages; see Appendix~\ref{sec:diarization_details}). We set $\lambda_{time}=5.0$ and $\lambda_{spk}=10.0$. We ensured all experiments started from the exact same seed, with identical data.

\textbf{Evaluation Datasets.} To comprehensively assess the diverse capabilities of \method{}, we employ a multi-faceted suite of evaluation benchmarks. We evaluate \textit{timestamped dense captioning} performance using the test set from TACOS~\cite{primus2025tacos}. To assess our \method$\rightarrow$LLM cascade, we leverage audio understanding \& reasoning benchmarks including MMAU~\cite{sakshi2024mmau}, MMAR~\cite{mmar}, MMSU~\cite{wang2025mmsu}, and MMAU-Pro~\cite{kumar2025mmau}. We evaluate our \method-V$\rightarrow$LLM cascade (Section~\ref{sec:av_pipeline}) on Daily-Omni~\cite{dailyomni}, World-Sense~\cite{worldsense}, Video-Holmes~\cite{holmes}, and AVHBench~\cite{avhbench}. For TACOS~\cite{primus2025tacos}, we adopt the evaluation metrics described in Section \ref{subsec:evaluation}, while for all other benchmarks we adopt their standard metrics.

\subsection{Dense Captioning}
\label{sec:results}

\newcommand{\hdrangle}{90}

\begin{table*}[t]
    \centering
    \begin{minipage}[t]{0.5\textwidth}
        \centering
        \setlength\tabcolsep{3pt}
        \scalebox{0.65}{
        \begin{tabular}{@{}l cccccc ccc ccccc@{}}
        & \rotatebox{\hdrangle}{Multitask} & \rotatebox{\hdrangle}{Pretrained} & \rotatebox{\hdrangle}{Templates} & \rotatebox{\hdrangle}{Acoustic Sim} & \rotatebox{\hdrangle}{TACOS} & \rotatebox{\hdrangle}{Diarization}
        & \rotatebox{\hdrangle}{Iters} & \rotatebox{\hdrangle}{LoRA} & \rotatebox{\hdrangle}{TS Wt} 
        & \rotatebox{\hdrangle}{\textbf{EvtF1}$\uparrow$} & \rotatebox{\hdrangle}{SegF1} & \rotatebox{\hdrangle}{\textbf{Hal\%}$\downarrow$} & \rotatebox{\hdrangle}{Conf} & \rotatebox{\hdrangle}{Spec} \\
        \midrule
        
        \textbf{Ours (TAC)} & \cmark & \cmark & \cmark & \cmark & \cmark & \xmark & 5k & 128 & 5.0 & \textbf{.50} & \textbf{.71} & 4.9 & 0.89 & 0.74 \\
        \textbf{Ours (TAC-Dia.)} & \cmark & \cmark & \cmark & \cmark & \cmark & \cmark & 5k & 128 & 5.0 & {.49} & \textbf{.71} & 3.0 & 0.91 & 0.77 \\
        \midrule
        \textit{Ablations} \\[-0.5ex]
        \quad \xmark~Multitask & \xmark & \gc & \gc & \gc & \gc & \xmark & \g{5k} & \g{128} & \g{5.0} & .45 & .72 & 7.0 & 0.87 & 0.70 \\
        \quad \hspace{1.5em}\textit{(merge=0.1)} & \xmark & \gc & \gc & \gc & \gc & \xmark & \g{5k} & \g{128} & \g{5.0} & .41 & .71 & 13.8 & 0.80 & 0.70 \\
        \quad \xmark~Pretrained & \gc & \xmark & \gc & \gc & \gc & \xmark & \g{5k} & \g{128} & \g{5.0} & .49 & .70 & 8.8 & 0.85 & 0.70 \\
        \quad \xmark~Templates & \gc & \gc & \xmark & \gc & \gc & \xmark & \g{5k} & \g{128} & \g{5.0} & .47 & .71 & 2.2 & 0.93 & 0.78 \\
        \quad \xmark~Acoustic Sim & \gc & \gc & \gc & \xmark & \gc & \xmark & \g{5k} & \g{128} & \g{5.0} & .49 & .71 & 5.3 & 0.89 & 0.75 \\
        \quad \xmark~TACOS & \gc & \gc & \gc & \gc & \xmark & \xmark & \g{5k} & \g{128} & \g{5.0} & .42 & .68 & 7.6 & 0.85 & 0.70 \\
        
        \midrule
        \multirow{3}{*}{\textit{LoRA Rank}} 
        & \multirow{3}{*}{\gc} & \multirow{3}{*}{\gc} & \multirow{3}{*}{\gc} & \multirow{3}{*}{\gc} & \multirow{3}{*}{\gc} & \xmark & \g{5k} & 256 & \g{5.0} & .48 & .70 & 3.5 & 0.90 & 0.75 \\
        & & & & & & \xmark & \g{5k} & 64 & \g{5.0} & .49 & .71 & 4.8 & 0.89 & 0.74 \\
        & & & & & & \xmark & \g{5k} & 8 & \g{5.0} & .19 & .66 & 36.0 & 0.58 & 0.54 \\
        
        \midrule
        \multirow{2}{*}{\textit{Timestamp weight}} 
        & \multirow{2}{*}{\gc} & \multirow{2}{*}{\gc} & \multirow{2}{*}{\gc} & \multirow{2}{*}{\gc} & \multirow{2}{*}{\gc} & \xmark & \g{5k} & \g{128} & 1.0 & .48 & .71 & 4.2 & 0.91 & 0.76 \\
        & & & & & & \xmark & \g{5k} & \g{128} & 10.0 & .48 & .71 & 5.8 & 0.88 & 0.73 \\
        
        \midrule
        \multirow{2}{*}{\textit{Iterations}} 
        & \multirow{2}{*}{\gc} & \multirow{2}{*}{\gc} & \multirow{2}{*}{\gc} & \multirow{2}{*}{\gc} & \multirow{2}{*}{\gc} & \xmark & 10k & \g{128} & \g{5.0} & .47 & .70 & 5.2 & 0.89 & 0.75 \\
        & & & & && \xmark  & 2.5k & \g{128} & \g{5.0} & .46 & .70 & 8.0 & 0.85 & 0.72 \\
        
        \midrule
        \multicolumn{14}{c}{\textit{Baselines}} \\  \midrule 
        \quad Gemini 3 Pro & \g{--}  \g{--} & \g{--} & \g{--} & \g{--} & \g{--} & \g{--} & \g{--} & \g{--} & .42 & .64 & 6.1 & 0.84 & 0.66 \\
        \quad Qwen3-Omni & \g{--}  \g{--} & \g{--} & \g{--} & \g{--} & \g{--} & \g{--} & \g{--} & \g{--} & .37 & .66 & 7.3 & 0.84 & 0.62 \\
        \quad Audio Flamingo 3 & \g{--}  \g{--} & \g{--} & \g{--} & \g{--} & \g{--} & \g{--} & \g{--} & \g{--} & .27 & .55 & 11.6 & 0.73 & 0.59 \\
        
        \bottomrule
        \end{tabular}
        }
        \subcaption{Training Ablations \& Baselines}
        \label{tab:ablation_results}
    \end{minipage}
    \hspace{3em}
    \begin{minipage}[t]{0.35\textwidth}
        \centering
        \setlength\tabcolsep{3pt}
        \resizebox{\linewidth}{!}{
        \begin{tabular}{@{}cccc ccccc@{}}
        \rotatebox{90}{Style} & \rotatebox{90}{Merge} & \rotatebox{90}{Activity} & \rotatebox{90}{Resolution} 
        & \rotatebox{90}{\textbf{EvtF1}$\uparrow$} & \rotatebox{90}{SegF1} & \rotatebox{90}{\textbf{Hal\%}$\downarrow$} & \rotatebox{90}{Conf} & \rotatebox{90}{Spec} \\
        \midrule
        
        \textbf{brief} & \textbf{0.25} & \textbf{0.05} & \textbf{0.10} & \textbf{.50} & \textbf{.71} & \textbf{4.9} & \textbf{0.89} & \textbf{0.74} \\
        
        \midrule
        detailed & \g{0.25} & \g{0.05} & \g{0.10} & .49 & .71 & 8.0 & 0.86 & 0.72 \\
        keywords & \g{0.25} & \g{0.05} & \g{0.10} & .47 & .66 & 1.3 & 0.89 & 0.78 \\
        
        \midrule
        \g{brief} & 0.10 & \g{0.05} & \g{0.10} & .31 & .66 & 20.2 & 0.73 & 0.67 \\
        \g{brief} & 0.50 & \g{0.05} & \g{0.10} & .48 & .72 & 4.0 & 0.90 & 0.74 \\
        \g{brief} & 1.00 & \g{0.05} & \g{0.10} & .42 & .72 & 4.7 & 0.89 & 0.69 \\
        
        \midrule
        \g{brief} & \g{0.25} & 0.01 & \g{0.10} & .49 & .72 & 4.7 & 0.89 & 0.74 \\
        \g{brief} & \g{0.25} & 0.10 & \g{0.10} & .49 & .70 & 5.5 & 0.88 & 0.76 \\
        \g{brief} & \g{0.25} & 0.20 & \g{0.10} & .45 & .70 & 4.5 & 0.90 & 0.76 \\
        
        \midrule
        \g{brief} & \g{0.25} & \g{0.05} & 0.01 & .43 & .71 & 11.8 & 0.83 & 0.73 \\
        \g{brief} & \g{0.25} & \g{0.05} & 0.50 & .48 & .70 & 5.4 & 0.88 & 0.77 \\
        
        \bottomrule
        \end{tabular}
        }
        \subcaption{Inference Parameter Sweeps}
        \label{tab:sweep_results}
    \end{minipage}
    \caption{\small\textbf{Comprehensive Evaluation.} (a) Training ablations and baselines. (b) Post-hoc controllability analysis on the TACOS test set; the bold default was fixed on a separate development set before evaluation and was not selected using this sweep.}
    \label{tab:all_results}
    \vspace{-2em}
\end{table*}

We evaluate \method{} on the held-out test set of the \textbf{TACOS} benchmark, and zero-shot performance on AudioCaps~\cite{kim2019audiocaps} and Clotho-v2~\cite{drossos2020clotho}, unlike baselines trained on these datasets (Table~\ref{tab:captioning_results}), using METEOR, CIDEr, SPICE, and FENSE~\cite{fense}. For TACOS, every baseline was evaluated both zero-shot and few-shot with in-context examples of the required output format, and we report whichever condition scored higher. This permissive selection was made on the test set for baselines, whereas TAC's configuration was fixed on a separate development set; the comparison is therefore optimistic toward the baselines. Timestamped captioning results are summarized in Table~\ref{tab:all_results}.

\label{subsec:main_results}

\textbf{Comparison with State-of-the-Art:}
We first analyze the bottom section of Table~\ref{tab:all_results}. \method{} achieves a new state-of-the-art across all major temporal and semantic metrics, significantly outperforming previous state-of-the art models. The most striking improvement is in temporal grounding. We observe that for Event F1 score (EvtF1) our \method{} beats Qwen 3 Omni by $0.14$ F1 Score, and Gemini 3 Pro by $0.08$ F1 Score. Outside of temporal grounding, \method{} also out-performs in text-audio similarity ($0.89$ vs $0.84$), and Segment F1 score ($0.71$ vs $0.66$ for Gemini 3 Pro / $0.64$ for Qwen3-Omni). Competing models perform decently at ``global'' recognition, but fail to localize events precisely in dense mixtures. Our approach yields the lowest Hallucination Rate (\textbf{4.9\%}), nearly half that of the open-source baseline Audio Flamingo 3 ($11.6\%$) and significantly lower than Gemini 3 Pro ($6.1\%$). Furthermore, our high Specificity score ($0.74$) indicates that \method{} does not merely ``spot'' keywords but accurately describes the full duration of acoustic events.

\textbf{Ablation study:} We conduct a thorough ablation study of \method{}, varying each component one by one and studying its impact on temporal grounding and semantic similarity as shown in Table~\ref{tab:ablation_results}. We observe that multitask prompts are critical: static tasks drop EvtF1 from $0.50$ to $0.45$, and a poor merge threshold ($0.1$s) causes severe degradation ($0.41$ EvtF1, $13.8\%$ Hal). Pretraining cuts hallucination in half ($8.8\%{\rightarrow}4.9\%$), while scene templates improve EvtF1 ($0.47{\rightarrow}0.50$) at the cost of slightly higher hallucination - without templates, the model becomes overly conservative. LoRA rank $128$ seemed to work best and rank $8$ caused collapse. Training beyond $5$k iterations also degraded performance due to overfitting. The timestamp loss weight $\lambda_{time}{=}5.0$ balances EvtF1 and hallucination, and removing TACOS causes a large drop ($0.42$ EvtF1), confirming that real-world annotations helps generalization.
\textbf{Prompt controllability:} Table~\ref{tab:sweep_results} is a post-hoc analysis of how user controls affect the fixed TAC checkpoint, not a test-set hyperparameter search. The default (brief, merge $0.25$s, activity $0.05$, resolution $0.10$s) was selected on a separate development set before evaluation and is used throughout. Moving merge to $0.1$s sharply lowers Event F1 and raises hallucination; keyword captions have low hallucination because they are conservative; and a higher activity threshold widens event spans, trading temporal recall for confidence.

\subsection{Speaker Diarization}
\label{subsec:diarization}

\begin{wraptable}{r}{0.5\textwidth}
    \centering
    \vspace{-3em}
    \resizebox{\linewidth}{!}{
    \begin{tabular}{@{}lcccccr@{}}
    \toprule
    \textbf{Model} & \textbf{VoxConv.} & \textbf{AMI} & \textbf{AISH.-4}  & \textbf{Bazinga} & \textbf{Avg.} & \textbf{K-acc\%} \\
    \midrule
    \textbf{\method{}-Dia. (ours)} & \underline{6.33} & {11.10} & \underline{6.81} & \textbf{28.29} & \textbf{10.16} & \underline{71.0} \\
    \quad\textit{w/o sp.\ count} & 6.37 & \textbf{10.74} & 6.92 & 29.75 & 10.37 & 69.7 \\ \midrule
    DiariZen & \textbf{5.91} & 22.51 & \textbf{5.70} & \underline{38.98} & \underline{12.60} & 68.5 \\
    Sortformer-strm. & 6.15 & 24.96 & 8.45 & 46.39 & 14.57 & \textbf{71.9}\\
    Pyannote c-1 & 7.55 & \underline{12.83} & 6.88 & 47.30 & 13.84 & 61.2\\
    Pyannote-3.0 & 7.47 & 13.48 & 7.06 & 48.00 & 14.00 & 59.6\\
    Pyannote-3.1 & 7.44 & 13.43 & 6.94 & 48.29 & 14.00 & 59.8\\
    Reverb-v2 & 16.33 & 36.25 & 13.74 & 150.96 & 37.80 & 54.0\\
    \bottomrule
    \end{tabular}}
    \caption{\small \textbf{Speaker Diarization} (DER\%$\downarrow$, strict collar=0.25s). K-acc = \% windows with exactly correct speaker count. \method{} first counts speakers, then diarizes, achieving the lowest pooled DER across all four datasets.}
    \label{tab:diarization}
    \vspace{-1em}
\end{wraptable}

Because \method{} produces timestamped speech events with precise boundaries, it can be directly fine-tuned for speaker diarization - the task of determining ``who spoke when.'' We fine-tune on a multi-corpus mixture of ${\sim}$943 raw hours (${\sim}$468 effective) from 10 corpora across 6 languages with per-family equalization (see Appendix~\ref{sec:diarization_details} for full details). At inference, \method{}-Dia. first counts speakers, then diarizes conditioned on the predicted count. We evaluate on four benchmarks using Diarization Error Rate (DER\%) with a strict collar of~0.25s. Table~\ref{tab:diarization} shows that \method{} achieves the lowest pooled DER across all four datasets, outperforming dedicated systems including pyannote-3.1~\cite{pyannote}, DiariZen~\citep{raghav2026diarizen}, and Sortformer~\citep{sortformer}. The count-then-diarize strategy consistently improves both DER and speaker-counting accuracy (K-acc) over direct diarization (see Appendix~\ref{sec:diarization_prompt} for the prompt and sample outputs).

\begin{table*}[t]
    \centering
    \begin{minipage}[t]{0.48\textwidth}
        \centering
        \resizebox{\linewidth}{!}{
        \begin{tabular}{@{}llcccc@{}}
        \toprule
        \textbf{Dataset} & \textbf{Model} & \textbf{METEOR} & \textbf{CIDEr} & \textbf{SPICE} & \textbf{FENSE} \\
        \midrule
        \multirow{4}{*}{Clotho-v2} & Audio Flamingo 3 & 13.74 & 43.94 & 11.28 & 52.1 \\
        & Qwen3-Omni-Captioner & 12.13 & 43.85 & 10.46 & 52.43 \\
        & Audio Flamingo Next & 18.58 & 46.45 & 13.84 & 56.21 \\
        & \textbf{TAC (ours, zero-shot)} & \textbf{19.07} & \textbf{48.83} & \textbf{14.07} & \textbf{58.4} \\
        \midrule
        \multirow{4}{*}{AudioCaps} & Audio Flamingo 3 & 23.28 & 70.04 & 16.56 & 58.2 \\
        & Qwen3-Omni-Captioner & 19.46 & 65.84 & 17.16 & 59.27 \\
        & Audio Flamingo Next & 24.39 & \textbf{71.16} & 16.92 & 62.4 \\
        & \textbf{TAC (ours, zero-shot)} & \textbf{24.53} & 68.29 & \textbf{17.87} & \textbf{63.1} \\
        \bottomrule
        \end{tabular}
        }
        \subcaption{Standard Audio Captioning}
        \label{tab:captioning_results}
    \end{minipage}
    \hfill
    \begin{minipage}[t]{0.48\textwidth}
        \centering
        \resizebox{\linewidth}{!}{
        \begin{tabular}{@{}l cc cc c@{}}
        \toprule
        & \multicolumn{2}{c}{\textbf{VDC Detailed}} & \multicolumn{2}{c}{\textbf{V-SAL~2 Test}} & \textbf{Omni-Cloze} \\
        \cmidrule(lr){2-3} \cmidrule(lr){4-5} \cmidrule(lr){6-6}
        \textbf{Model} & \%Acc$\uparrow$ & Score$\uparrow$ & \%Miss$\downarrow$ & \%Hall$\downarrow$ & Total$\uparrow$ \\
        \midrule
        GPT-4o & 46.3 & 2.5 & 17.0 & 14.2 & -- \\
        Gemini 1.5 Pro & 43.1 & 2.2 & 21.8 & 16.5 & -- \\
        Qwen2.5-Omni-7B & 39.7 & 2.2 & 26.3 & 21.7 & -- \\
        Video-LLaMA~2 & 29.7 & 1.4 & 56.8 & 8.9 & 6.6 \\
        Video-SALMONN~2 & 46.1 & 2.5 & 10.0 & 12.9 & 3.3 \\
        Omni-Captioner & 55.0 & 2.7 & 17.8 & 10.9 & 56.4 \\
        Gemini 3.1 Pro & 56.4 & 3.5 & 12.2 & 7.4 & -- \\
        \textbf{TAC-V (ours)} & \textbf{59.8} & \textbf{3.6} & \textbf{8.7} & \textbf{5.2} & \textbf{61.9} \\
        \bottomrule
        \end{tabular}
        }
        \subcaption{Dense Audiovisual Captioning}
        \label{tab:dense_captioning}
    \end{minipage}
    \caption{\small\textbf{Captioning Evaluation.} (a) Standard audio captioning on Clotho-v2 and AudioCaps. (b) Dense audiovisual captioning on VDC~\cite{auroracap}, Video-SALMONN~2 Test Set~\cite{videosalmonn2}, and Omni-Cloze~\cite{omnicaptioner} (Total = avg. of Visual, Audio, and AV subsets).}
    \label{tab:captioning_combined}
    \vspace{-1em}
\end{table*}

\section{Describe-Then-Reason}
\label{sec:semantic_bridge}


We now turn to using \method{} and its audiovisual extension \method{}-V as a semantic bridge to a text-only reasoner. Here, \method{}(-V) converts audio or video into a precise timestamped text representation; the reasoner never sees the original media. We call this paradigm ``describe-then-reason''. We evaluate Qwen3-8B, Qwen3-Next-80B-A3B-Thinking~\cite{qwen3technicalreport}, and Gemini 3 Pro~\cite{gemini25}. Scaling helps while reasoning is the bottleneck, but can saturate when the caption omits task-specific evidence. The small-reasoner and matched-captioner controls in Appendix~\ref{sec:rebuttal_results} separate caption quality from reasoner capacity.

\subsection{Audio Understanding \& Reasoning}
\label{subsec:audio_reasoning}

Table~\ref{tab:audio_results} shows results on complex audio understanding and reasoning benchmarks. With Qwen3-Next-80B, \method{} reaches $73.9\%$ on MMAU; Gemini 3 Pro yields $71.9\%$ on MMAR, $72.4\%$ on MMSU, and $62.9\%$ on MMAU-Pro. A much smaller Qwen3-8B still reaches $64.2/51.2/61.3$ on MMAU/MMAR/MMSU. A dedicated-captioner control, Qwen3-Omni-Captioner$\rightarrow$Gemini 3.1 Pro, obtains $65.4/67.9/57.1$ on MMAU-mini/MMAU/MMAR, below TAC captions with Gemini 3 Pro ($72.2/72.2/71.9$). Thus both caption quality and reasoner capacity matter (Appendix~\ref{sec:rebuttal_results}).
\begin{table*}[t]
    \centering
    \begin{minipage}[t]{0.43\textwidth}
        \centering
        \resizebox{\linewidth}{!}{
        \begin{tabular}{ll c | cc}
        \toprule
        & \multicolumn{2}{c|}{\textbf{Prior Methods}} & \multicolumn{2}{c}{\textbf{TAC + Text-only Reasoner}} \\
        \cmidrule(lr){2-3} \cmidrule(lr){4-5}
        \textbf{Benchmark} & Model & Score & + Qwen3 & + Gemini3 \\
        \midrule
        MMAU & Audio Thinker & \textbf{75.9} & 73.9 & 72.2 \\
        \quad \textit{Sound} & & 78.8 & \textbf{79.7} & 79.6 \\
        \quad \textit{Music} & & \textbf{73.8} & 62.6 & 63.4 \\
        \quad \textit{Speech} & & 75.2 & \textbf{79.3} & 73.6 \\
        \midrule
        \multirow{2}{*}{MMAU-mini} & Gem. Flash$\rightarrow$Gem. 3 Pro & 68.5 & \multirow{2}{*}{71.7} & \multirow{2}{*}{\textbf{72.2}} \\
        & Gemini 2.5 Flash & 64.3 & & \\ \midrule
        \multirow{2}{*}{MMAR} & Gem. Flash$\rightarrow$Gem. 3 Pro & 65.9 & \multirow{2}{*}{60.1} & \multirow{2}{*}{\textbf{71.9}} \\
        & Audio Flamingo 3 & 60.1 & & \\ \midrule
        MMSU & Audio Flamingo 3 & 62.3 & 65.0 & \textbf{72.4} \\
        MMAU-Pro & Gemini 2.5 Flash & 59.2 & 62.5 & \textbf{62.9} \\
        \bottomrule
        \end{tabular}
        }
        \subcaption{Audio Understanding \& Reasoning}
        \label{tab:audio_results}
    \end{minipage}
    \hspace{0.5em}
    \begin{minipage}[t]{0.54\textwidth}
        \centering
        \resizebox{\linewidth}{!}{
        \begin{tabular}{ll c | cc ccc}
        \toprule
        & \multicolumn{2}{c|}{\textbf{Prior Methods}} & \multicolumn{4}{c}{\textbf{Describer + Text-only Reasoner}} \\
        \cmidrule(lr){2-3} \cmidrule(lr){4-8}
        \textbf{Benchmark} & Model & Score & \shortstack{VLM \\ + Qwen3} & \shortstack{V-LLaMA2 \\ + Gem3} & \shortstack{V-SAL2 \\ + Gem3} & \shortstack{TAC-V \\ + Qwen3} & \shortstack{TAC-V \\ + Gem3} \\
        \midrule
        \multirow{3}{*}{Daily-Omni} & Qwen3-Omni & 76.2 & \multirow{3}{*}{51.5} & \multirow{3}{*}{41.4} & \multirow{3}{*}{68.4} & \multirow{3}{*}{{72.9}} & \multirow{3}{*}{\textbf{77.9}} \\
        & Gemini 2.5 Flash & 72.7 & & & & \\
        & OmniVinci & 66.5 & & & & \\
        \midrule
        \multirow{2}{*}{World-Sense} & Gemini 2.5 Pro & \textbf{65.1} & \multirow{2}{*}{37.4} & \multirow{2}{*}{40.7} & \multirow{2}{*}{42.2} & \multirow{2}{*}{{45.7}} & \multirow{2}{*}{58.6} \\
        & OmniVinci & 48.2 & & & & \\
        \midrule
        Video-Holmes & Qwen3-Omni & 57.3 & 45.6 & 39.3 & 47.6 & \multirow{1}{*}{{47.7}} & \textbf{59.2} \\
        \midrule
        AVHBench (AVH) & PandaGPT & 58.5 & 70.8 & 70.4 & 71.2 & \multirow{1}{*}{{79.8}} & \textbf{81.7} \\
        AVHBench (VAH) & PandaGPT & 61.3 & 51.8 & 46.2 & 41.4 & \multirow{1}{*}{{76.1}} & \textbf{76.6} \\
        AVHBench (AVM) & OneLLM & 60.1 & 50.5 & 53.1 & 55.7 & \multirow{1}{*}{{56.7}} & \textbf{61.6} \\
        AVHBench (AVC) & Video-LLaMA~2 & 14.0 & 12.9 & 7.4 & 6.9 & \multirow{1}{*}{\textbf{22.6}} & {20.6} \\
        \bottomrule
        \end{tabular}
        }
        \subcaption{Audio-visual Understanding \& Reasoning}
        \label{tab:av_results}
    \end{minipage}
    \caption{\small\textbf{Downstream Reasoning Benchmarks.} We compare prior SOTA end-to-end and cascade methods against our cascade approach: TAC/TAC-V captions fed to a text-only reasoner. We include cascade baselines (e.g., Gemini~Flash$\rightarrow$Gemini~3~Pro, Video-LLaMA~2/Video-SALMONN~2$\rightarrow$Gemini~3~Pro) for fair comparison using the same downstream reasoner.}
    \label{tab:combined_results}
\end{table*}

\subsection{Audiovisual Understanding \& Reasoning}
\label{subsec:av_results_analysis}

We apply \method{}-V (Sec. \ref{sec:av_pipeline}) to obtain dense timestamped audiovisual captions. We evaluate the quality of our generated audiovisual captions by using them as the \textit{sole input} for downstream reasoning tasks. In this setup, the reasoning Large Language Model (LLM) \textit{sees no video or audio}; it must answer complex questions based entirely on the text description generated by \method{}-V. Table~\ref{tab:av_results} presents results against SOTA native multimodal models. Our text-based cascade achieves SOTA on \textit{Daily-Omni} and \textit{Video-Holmes}, and significant gains on \textit{AVHBench}, which explicitly measures cross-modal hallucination. By separating event detection from visual grounding, our pipeline regularizes against the hallucinations common in end-to-end models (see Section~\ref{sec:qual_av} for examples). Crucially, alternative cascades using the same Gemini~3~Pro reasoner but weaker captioners all underperform \method{}-V, confirming that the gains stem from our caption quality (Table~\ref{tab:av_results}). We additionally evaluate on dedicated dense captioning benchmarks (Table~\ref{tab:dense_captioning}), where \method{}-V surpasses the prior SOTA Omni-Captioner~\cite{omnicaptioner} on VDC Detailed~\cite{auroracap} and Omni-Cloze, and achieves the lowest miss and hallucination rates on the Video-SALMONN~2 Test Set~\cite{videosalmonn2}.

\section{Conclusion, Limitations, and Future Work}
\label{sec:conclusion}

In this work, we introduced \method{}, a model that bridges the gap between raw acoustic signals and high-level reasoning through temporal dense captioning. We showed that robust temporal grounding can be learned from purely synthetic mixtures. We further extended \method{} with a VLM, producing \method{}-V, which generates rich, high-quality dense audio-visual captions. \method{} achieves state-of-the-art performance on dense captioning benchmarks--surpassing frontier systems such as Gemini 3 Pro. When cascaded with text-only LLMs, both TAC and TAC-V serve as powerful semantic bridges for downstream reasoning, unlocking expert-level state-of-the-art performance on audio and audio-visual reasoning benchmarks, respectively.

Despite these advancements, synthetic training can cause domain shift: the model sometimes over-predicts dramatic events (e.g., gunshots) in mundane videos and lacks fine-grained musical precision (e.g., chord progressions). The mixer partly targets this risk with convolution over 20.6K measured room impulse responses, sparse room-tone scenes, and templates spanning high-noise, short-segment, and event-density extremes. All evaluations use real recordings; zero-shot AudioCaps/Clotho and the held-out MSDWild results provide encouraging evidence, but do not cover all environments. Future work will study unsupervised domain adaptation and calibration of event priors. Describe-then-reason is also reusable: one caption can answer many questions without reprocessing the media, though performance saturates when the caption omits the evidence a stronger reasoner needs.

\bibliographystyle{plainnat}
\bibliography{references}


\newpage
\appendix
\section{Appendix}
\begin{itemize}
    \item Section~\ref{sec:impact}: Impact Statement
    \item Section~\ref{sec:baselines}: Baselines
    \item Section~\ref{sec:dataset_details}: Dataset \& Pipeline Details
    \item Section~\ref{sec:qual_audio}: Qualitative Analysis: Audio Understanding \& Reasoning
    \item Section~\ref{sec:qual_av}: Qualitative Analysis: Audio-Visual Understanding
    \item Section~\ref{sec:prompts}: Prompts
    \item Section~\ref{sec:llmusage}: LLM Usage
    \item Section~\ref{sec:rebuttal_results}: Additional Validation and Efficiency Results
\end{itemize}

\section{Additional Validation and Efficiency Results}
\label{sec:rebuttal_results}

\paragraph{Human validation and threshold sensitivity.}
For 100 sampled TACOS predictions, human raters judged whether every predicted event was present in the audio. Both automatic measures correlate strongly with these judgments: Spearman/Kendall correlations are $0.81/0.76$ for the LLM semantic judge and $0.73/0.69$ for FLAM confidence. Table~\ref{tab:flam_sensitivity} varies the FLAM threshold without changing predictions. TAC has the lowest Hal\% from $\tau=0.20$ upward. At $\tau=0.15$, TAC and Gemini 3 Pro are effectively tied; this rate should be read with recall because Gemini is more conservative (EvtF1/SegF1 $0.42/0.64$ versus TAC's $0.50/0.71$).

\begin{table}[h]
\centering
\caption{Hallucination rate (Hal\%$\downarrow$) across FLAM thresholds on TACOS.}
\label{tab:flam_sensitivity}
\setlength\tabcolsep{5pt}
\begin{tabular}{lrrrrrr}
\toprule
$\tau$ & .15 & .20 & .25 & .30 & .40 & .50 \\
\midrule
TAC (ours) & 3.8 & \textbf{4.3} & \textbf{4.9} & \textbf{5.5} & \textbf{7.0} & \textbf{8.3} \\
Gemini 3 Pro & \textbf{3.7} & 5.1 & 6.1 & 7.2 & 9.6 & 12.3 \\
Audio Flamingo 3 & 8.9 & 10.2 & 11.6 & 13.1 & 16.7 & 21.6 \\
\bottomrule
\end{tabular}
\end{table}

\paragraph{Cross-corpus diarization.}
We evaluate zero-shot on MSDWild, which is absent from the training mixture, using 30-second windows. TAC-Dia improves DER on both splits: $17.80/42.10$ (few/many speakers), compared with pyannote's $21.96/43.15$ and DiariZen's $22.42/45.70$.

\paragraph{Reasoner and captioner controls.}
Table~\ref{tab:reasoner_controls} reports the discussion-period controls. Holding TAC captions fixed, moving from Qwen3-8B to Qwen3-Next-80B substantially improves all three benchmarks; Gemini 3 Pro further improves reasoning-heavy MMAR and MMSU but is flat on MMAU. Conversely, even with the stronger Gemini 3.1 Pro reasoner, the untimestamped Qwen3-Omni-Captioner cascade trails TAC$\rightarrow$Gemini 3 Pro.

\begin{table}[h]
\centering
\caption{Describe-then-reason controls (accuracy, \%). A dash denotes an unreported result.}
\label{tab:reasoner_controls}
\setlength\tabcolsep{4pt}
\begin{tabular}{lrrrr}
\toprule
Describer $\rightarrow$ reasoner & MMAU & MMAU-mini & MMAR & MMSU \\
\midrule
Qwen3-Omni-Captioner $\rightarrow$ Gemini 3.1 Pro & 67.9 & 65.4 & 57.1 & -- \\
TAC $\rightarrow$ Qwen3-8B & 64.2 & 65.1 & 51.2 & 61.3 \\
TAC $\rightarrow$ Qwen3-Next-80B & \textbf{73.9} & 71.7 & 60.1 & 65.0 \\
TAC $\rightarrow$ Gemini 3 Pro & 72.2 & \textbf{72.2} & \textbf{71.9} & \textbf{72.4} \\
\bottomrule
\end{tabular}
\end{table}

\paragraph{Latency.}
Table~\ref{tab:latency} reports wall-clock seconds on one A100. TAC is served with vLLM; the VLM captions 30-second chunks at 2 fps; Whisper large-v3-turbo processes only detected speech; API network time is included for the reasoner. TAC accounts for under 10\% of the 10-minute end-to-end total. The caption is computed once and can be reused for multiple questions.

\begin{table}[h]
\centering
\caption{Pipeline latency in seconds.}
\label{tab:latency}
\begin{tabular}{lrrrr}
\toprule
Input & TAC & Whisper & VLM (TAC-V) & Gemini 3.1 Pro \\
\midrule
30 s & 1.4 & 1.3 & 15.0 & 20.0 \\
1 min & 2.6 & 2.9 & 34.0 & 35.0 \\
10 min & 39.2 & 31.6 & 256.0 & 188.0 \\
\bottomrule
\end{tabular}
\end{table}

\paragraph{Reproducibility.}
The architecture, optimization settings, data-stream proportions, prompts, metric definitions, judge rubric, matching rule, and scene-generation procedure are specified in the paper and appendix. The timestamped supervision is generated algorithmically rather than by the LLM used for continued-pretraining captions. Licensed source libraries cannot be redistributed, but the mixer is source-pool agnostic and can be reconstructed with freely licensed audio; the backbone, LoRA adaptation, FLAM, and Whisper are publicly available components. Because this work was conducted in an industrial setting, release of code and trained weights remains subject to internal and legal review; we therefore make no release commitment here.

\paragraph{In-the-wild evidence and remaining domain shift.}
All quantitative evaluations use real recordings, including TACOS, AudioCaps, Clotho-v2, the multilingual diarization corpora, MSDWild, the audio reasoning benchmarks, and the audiovisual benchmarks. As a qualitative out-of-domain check, the project demo includes movie trailers with rapid cuts, layered sound design, music, effects, and overlapping dialogue; TAC remains temporally precise on these examples despite synthetic-mixture training. This evidence is encouraging but not exhaustive. Failure regimes include over-predicting dramatic events in mundane scenes, inseparable overlapping speech, highly unusual acoustics, and musical details such as chord progressions. Future evaluation should stratify real recordings by reverberation, sparsity, SNR, and event density.

\section{Impact Statement}
\label{sec:impact}
This work advances the reliability of Large Audio Language Models by significantly reducing hallucination rates, creating a pathway toward trustworthy AI for safety-critical monitoring and accessibility tools for the hearing impaired. While \method{} enables detailed, time-synchronized narratives that surpass coarse global captions, the ability to detect fine-grained events carries potential surveillance risks if misused for unauthorized analysis of private environments. Furthermore, while our synthetic mixing approach mitigates privacy leaks associated with uncurated web data, synthetic pipelines may still inherit biases from their source libraries. We encourage the community to adopt these robust supervision methods while developing safeguards to ensure equitable and privacy-preserving deployment.

\section{Baselines}
\label{sec:baselines}

\textbf{Dense captioning \& SED.} We compare \method{} against SOTA open-source, open-weights, and frontier baselines: Gemini 3 Pro~\cite{gemini25}, Qwen3-Omni-30B-A3B-Instruct~\cite{Qwen3Omni}, and Audio Flamingo 3~\cite{af3}. We use the Instruct variant of Qwen3-Omni as the captioner variant cannot accept task-specific instructions (e.g., multiple-choice questions), making it unsuitable for evaluation on understanding and reasoning benchmarks.

\textbf{Standard audio captioning.} For AudioCaps and Clotho-v2 (Table~\ref{tab:captioning_results}), we compare zero-shot TAC against Audio Flamingo 3~\cite{af3}, Qwen3-Omni-Captioner~\cite{Qwen3Omni}, and Audio Flamingo Next~\cite{afnext}. The baselines are trained on the target captioning datasets, whereas TAC is not.

\textbf{Dense audiovisual captioning.} For VDC, Video-SALMONN~2 Test, and Omni-Cloze (Table~\ref{tab:dense_captioning}), we compare \method{}-V against GPT-4o, Gemini 1.5 Pro and Gemini 3.1 Pro~\cite{gemini25}, Qwen2.5-Omni-7B, Video-LLaMA~2~\cite{videollama2}, Video-SALMONN~2~\cite{videosalmonn2}, and Omni-Captioner~\cite{omnicaptioner}. Published comparison values are taken from the Omni-Captioner evaluation protocol~\cite{omnicaptioner}. We could not obtain Qwen3.5-Omni access during the evaluation period and therefore do not report a result for it.

\textbf{Downstream reasoning (audio).} We compare our \method{}$\rightarrow$LLM cascade against Audio Thinker~\cite{thinker}, Audio Flamingo 3~\cite{af3}, Gemini 2.5 Flash~\cite{gemini25}, and published captioner cascades from Omni-Captioner~\cite{omnicaptioner}. For fair cascade comparison, we include Gemini~2.5~Flash$\rightarrow$Gemini~3~Pro and Qwen3-Omni-Captioner$\rightarrow$Gemini~3.1~Pro controls, as well as TAC with 8B, 80B, and frontier reasoners.

\textbf{Downstream reasoning (audiovisual).} We compare our \method{}-V$\rightarrow$LLM cascade against Qwen3-Omni~\cite{Qwen3Omni}, Gemini 2.5 Flash/Pro~\cite{gemini25}, Omni-Vinci~\cite{omnivinci}, PandaGPT~\cite{pandagpt}, OneLLM~\cite{onellm}, and Video-LLaMA~2~\cite{videollama2}. We also include cascade baselines using Video-LLaMA~2 and Video-SALMONN~2 as captioners with the same Gemini~3~Pro reasoner.

\textbf{Speaker diarization.} We compare against dedicated diarization systems: pyannote-3.0 and pyannote-3.1~\cite{pyannote}, pyannote c-1 (a community variant), DiariZen~\citep{raghav2026diarizen}, Sortformer~\citep{sortformer}, and Reverb-v2. All baselines are evaluated with DER\% at a strict collar of 0.25s.

\section{Dataset \& Pipeline Details}
\label{sec:dataset_details}

Table~\ref{tab:dataset_comparison} compares our synthetic pipeline against existing temporally grounded audio-caption resources. Our pipeline generates ${\sim}$150K synthetic mixtures (${\sim}$1.2K hours), with source audio drawn from ${\sim}$450 hours of music (${\sim}$18s avg), ${\sim}$290 hours of speech (${\sim}$4s per utterance), ${\sim}$150 hours of sound effects (${\sim}$3s per event), and ${\sim}$300 hours of background ambience (${\sim}$18s avg). Average caption lengths are 20 words (keyword style), 40 words (brief), and 55 words (detailed).

\subsection{Scene Templates, Source Pools, and Augmentation}
\label{sec:scene_template_details}
The six scene templates are handwritten rather than learned or LLM-generated and are sampled equally during training. Each template binds temporal constraints to roles drawn from ambience, speech, sound effects, transitional elements, music, and impulse responses. Unless otherwise stated, scene duration is sampled uniformly from 8--30 seconds and roles are sampled independently. Table~\ref{tab:scene_templates} reports every numeric distribution disclosed for the templates; ``as above'' means the same auxiliary ambience, music, transition, and effects roles as the Single Speaker template.

\begin{table}[h]
\centering
\caption{Scene-template sampling parameters. Within each role, entries give duration in seconds, count, appearance probability, and loudness in dB when reported.}
\label{tab:scene_templates}
\small
\setlength{\tabcolsep}{4pt}
\begin{tabular}{p{0.20\linewidth}p{0.74\linewidth}}
\toprule
\textbf{Template} & \textbf{Role distributions and constraints} \\
\midrule
Effects over Ambience & ambience $(18$--$30,1,0.8,-28$--$-23)$; effects $(1$--$5,1$--$8,1.0,-19$--$-17)$; music $(3$--$7,2,0.7,-24$--$-16)$ \\
Single Speaker & ambience $(10$--$30,1,0.9)$; speech $(4$--$30,1,1.0,-12$--$-8)$; music $(5$--$30,1,0.7)$; transitions $(1$--$30,1$--$4,0.8)$; effects $(4$--$7,1$--$8,1.0)$ \\
Overlapping Speakers & scene duration $12$--$30$; speech $(3$--$6,2$--$5,1.0,-12$--$-8)$; auxiliary roles as above \\
Speech over Music & ambience $(10$--$30,1,0.8,-23$--$-18)$; speech $(4$--$30,1$--$3,0.65,-19$--$-10)$; music $(5$--$30,1,0.9)$; effects $(1$--$4,1,1.0)$ \\
Loud Background & ambience fixed at $-12$ dB; speech, music, and effects at $-20$--$-16$ dB, producing negative SNR \\
Very Short Clips & scene duration $1$--$4$; ambience, speech, and music roles only \\
\bottomrule
\end{tabular}
\end{table}

The source pools contain approximately 1.1M sound-effect clips, 452.6K music clips, 644.8K speech clips, 2K room tones, and 20.6K measured impulse responses. These sources come from licensed commercial sound-effect, music, and speech libraries and therefore cannot be redistributed. Public real-world components include TACOS, AudioSet-Strong, and the ten diarization corpora in Appendix~\ref{sec:diarization_details}; the mixer is source-pool agnostic and can instead use freely licensed sources such as Freesound.

Signal-level augmentation is sampled independently per mixture. EQ tilt in $[-8,+8]$ dB, dynamic volume with 1--3 breakpoints in $[-35,-10]$ dB, and reverberation by convolution with a measured impulse response are each applied with probability $0.25$. Audio artifacts are injected with probability $0.2$. Roles additionally support dropout and retrigger controls.

\begin{table}[h]
\centering
\caption{Comparison with existing temporally grounded audio-caption resources. Avg.~Ev./Clip = average number of events per clip. Overlap = whether overlapping events are present. R/S = Real or Synthetic.}
\label{tab:dataset_comparison}
\vspace{0.5em}
\scalebox{0.58}{
\begin{tabular}{@{}lrc rrc lll@{}}
\toprule
\textbf{Dataset} & \textbf{Clips} & \textbf{Duration} & \textbf{Avg.~Dur.} & \textbf{Avg.~Ev./Clip} & \textbf{Overlap} & \textbf{Temporal Grounding} & \textbf{Caption Type} & \textbf{R/S} \\
\midrule
\textbf{TAC (ours)} & ${\sim}$150K & ${\sim}$1.2K hrs & ${\sim}$20s & ${\sim}$5 & \cmark & Event-level timestamps & Dense, open-vocab & Synthetic \\
TACOS~\cite{primus2025tacos} & 12,358 & 77 hrs & ${\sim}$22s & ${\sim}$3 & \cmark & Event-level timestamps & Dense, open-vocab & Real \\
AudioSet-Strong~\cite{audiosetstrong} & ${\sim}$288K & ${\sim}$334 hrs & ${\sim}$10s & ${\sim}$2 & \cmark & Event-level timestamps & Closed-set (527 classes) & Real \\
WavCaps~\cite{mei2024wavcaps} & ${\sim}$400K & ${\sim}$760 hrs & ${\sim}$7s & 1 & \xmark & None (clip-level) & Single caption (noisy) & Real \\
AudioCaps~\cite{kim2019audiocaps} & 46K & ${\sim}$128 hrs & ${\sim}$10s & 1 & \xmark & None (clip-level) & Single global caption & Real \\
Clotho~\cite{drossos2020clotho} & 6K & ${\sim}$24 hrs & ${\sim}$15s & 1 & \xmark & None (clip-level) & Single global caption & Real \\
\bottomrule
\end{tabular}
}
\end{table}

Algorithm~\ref{alg:scene_compilation} details the full dynamic scene mixing and supervision generation procedure used by our pipeline.

\begin{algorithm}[h]
   \caption{Dynamic Scene Mixing \& Supervision}
   \label{alg:scene_compilation}
   \small
\begin{algorithmic}
   \STATE {\bfseries Input:} Template $T$, Audio Sources $S$, Dynamic Params $\Theta_{dyn}$: Merge Threshold $\delta_{merge}$  Activity Threshold $\delta_{act}$, Resolution Threshold $\delta_{res}$
   \STATE {\bfseries Output:} Mixed Audio $A_{mix}$, Hierarchical Prompt $P$, Caption $Y$

   \STATE $E \leftarrow \text{InstantiateEvents}(T, S)$
   \STATE $A_{mix} \leftarrow \mathbf{0}$
   \FOR{each event $e_i \in E$}
       \STATE $a_i \leftarrow \text{ProcessAudio}(e_i)$ \COMMENT{Simulate reverb, fading, dist}
       \STATE $A_{mix} \leftarrow A_{mix} + a_i$
       \STATE $M_i \leftarrow \text{ComputeRMS}(a_i)$
   \ENDFOR

   \STATE \COMMENT{Dynamic Ground Truth Generation}
   \STATE $\delta_{merge}, \delta_{act}, \delta_{res} \sim \Theta_{dyn}$ \COMMENT{Sample supervision strictness}
   \STATE $Y \leftarrow []$
   \FOR{each event $e_i$}
       \STATE $R_i \leftarrow \text{GetNonZeroRanges}(M_i, \delta_{merge}, \delta_{act})$
       \STATE $L_i \leftarrow \text{GetLevel}(e_i, \text{style} \sim \{\text{brief, detailed, kw}\})$
       \STATE $Y.\text{append}(\text{Format}(L_i, R_i, \delta_{res}))$
   \ENDFOR
   \STATE $P \leftarrow \text{ConstructPrompt}(\Theta_{dyn}, \text{style})$
   \STATE \textbf{return} $A_{mix}, P, Y$
\end{algorithmic}
\end{algorithm}

\subsection{Diarization Training Details}
\label{sec:diarization_details}

As described in Section~\ref{sec:experiments}, \method{} is trained jointly on multiple data streams, one of which is real-world diarization data. The diarization stream accounts for 50\% of training ranks and draws from a multi-corpus mixture totaling ${\sim}$943 raw hours from 10 corpus families across 6 languages. The remaining ranks are allocated to SED data (25\%), the TAC synthetic mixer (12.5\%), and TACOS (12.5\%).

Table~\ref{tab:diarization_data} summarizes the diarization data composition. To prevent large corpora from dominating, we apply per-family equalization: each corpus family receives equal sampling probability, with video-with-faces sources (WASD, AVA-AVD, Bazinga, AMI-far) additionally downweighted to $0.25\times$. This yields 6 families at $1.0\times$ weight and 4 at $0.25\times$, for a total effective weight of $7.0$. The speaker-ID loss weight $\lambda_{spk}=10.0$ (Eq.~\ref{eq:weight_time_loss}) ensures the model learns robust speaker discrimination. We additionally apply caption-stream dropout with probability $0.3$ and set the diarization emission probability to $1.0$. Speech source descriptions are generated via Qwen3-VL captions on ${\sim}$17k voice clips.

\begin{table}[h]
\centering
\caption{Diarization training data composition. Weight refers to per-family sampling weight; Samp\% is effective sampling share after equalization.}
\label{tab:diarization_data}
\vspace{0.5em}
\scalebox{0.65}{
\begin{tabular}{@{}lrrcccl@{}}
\toprule
\textbf{Family} & \textbf{Raw hrs} & \textbf{Eff hrs} & \textbf{Weight} & \textbf{Avg spk} & \textbf{Samp\%} & \textbf{Langs} \\
\midrule
Bazinga & 493.1 & 123.3 & 0.25 & 16.5 & 3.6\% & en \\
CallHome & 99.1 & 99.1 & 1.00 & 2.4 & 14.3\% & de, ja, zh, en, es \\
AMI-IHM & 90.3 & 90.3 & 1.00 & 4.0 & 14.3\% & en \\
AISHELL-4 & 74.2 & 74.2 & 1.00 & 4.8 & 14.3\% & zh \\
CallFriend & 23.6 & 23.6 & 1.00 & 2.2 & 14.3\% & en, fr, ja \\
AMI-far & 89.3 & 22.3 & 0.25 & 4.0 & 3.6\% & en \\
VoxConverse & 18.8 & 18.8 & 1.00 & 4.9 & 14.3\% & en \\
AVA-AVD & 27.4 & 6.9 & 0.25 & 7.7 & 3.6\% & en \\
WASD & 23.5 & 5.9 & 0.25 & 2.2 & 3.6\% & en \\
SIMSAMU & 3.2 & 3.2 & 1.00 & 2.0 & 14.3\% & fr \\
\midrule
\textbf{Total} & 942.5 & 467.5 & -- & -- & 100\% & 6 langs \\
\bottomrule
\end{tabular}
}
\end{table}

\section{Qualitative Analysis: Audio Understanding \& Reasoning}
\label{sec:qual_audio}

In this section, we analyze the reasoning capabilities of the \method{}$\rightarrow$LLM cascade on purely acoustic tasks. A key advantage of our approach is the ability to perform deductive reasoning over the dense event logs generated by \method{}. Unlike end-to-end models that often output a direct answer, our pipeline generates an explicit "Thinking Trace" based on the timestamped captions, allowing for interpretability.

We present examples from the \textbf{MMAU-Pro} and \textbf{MMSU} benchmarks below.

\begin{figure}[h!]
\centering
\begin{tcolorbox}[colback=green!5!white, colframe=green!75!black, title=\textbf{MMAU-Pro: Complex Event Deduction}]
\small
\textbf{Question:} What is being prepared in the audio? \\
\textbf{Choices:} [A] None of these options, [B] Milk coffee, [C] Natural fruit juice, [D] Boba tea

\tcblower
\textbf{\method{} Caption Output:}
\textit{[0.2s - 5.9s] [background] Kitchen sounds with water (1.00)2. [  4.9s -  12.5s] [sfx] Sharp slicing sound (0.96)3. [  6.3s -  18.1s] [background] Kitchen sounds with water (1.00)4. [ 10.7s -  14.9s] [sfx] Metal can being opened (0.89)5. [ 16.8s -  17.0s] [sfx] Water drops into glass (1.00)6. [ 18.7s -  19.8s] [sfx] Sharp crack of wood (0.92)7. [ 20.2s -  23.4s] [background] Water running in sink (0.85)8. [ 23.2s -  23.4s] [sfx] Male breath sound (0.75)9. [ 23.8s -  30.2s] [background] Water running in sink (0.91)10. [ 25.7s -  26.0s] [sfx] Metal impact sound (0.99)11. [ 26.8s -  30.2s] [sfx] Water boiling in pot (1.00)12. [ 30.2s -  31.6s] [background] Distant traffic and birds (0.00)13. [ 31.8s -  33.4s] [sfx] Metal spoon tapping cup (1.00)14. [ 33.5s -  36.6s] [sfx] Pouring water into glass (0.99)15. [ 34.3s -  40.4s] [background] Distant traffic and birds (0.04)16. [ 39.1s -  39.3s] [sfx] Clicking and chewing (0.17)}

\vspace{0.1cm}
\textbf{LLM Reasoning Trace:}
Based on the provided audio caption, let's analyze the sequence of events to determine the answer:

1.  **Preparation sounds:** There are "kitchen sounds with water" and a "sharp slicing sound," indicating ingredients are being cut.
2.  **Specific container:** A "metal can being opened" is detected. This is a very specific sound often associated with sweetened condensed milk or evaporated milk, commonly used in certain beverages.
3.  **Liquids:** There are sounds of "water drops into glass," "water running in sink," "water boiling in pot," and "pouring water into glass."
4.  **Mixing:** A "metal spoon tapping cup" suggests stirring or mixing.
5.  **Deduction:** While slicing could imply fruit (juice), the combination of a *metal can being opened*, *boiling water*, and *spoon tapping a cup* strongly points towards a hot beverage preparation that involves canned milk, such as coffee with condensed milk. "Natural fruit juice" typically involves blending or squeezing, not boiling water or opening cans (unless it's canned juice, but the slicing suggests fresh ingredients, making the can the outlier for juice). Boba tea usually involves shaking ice, not necessarily the specific "spoon tapping cup" sound of stirring a hot drink, though possible. However, the classic sequence of slicing (perhaps a lemon or just general prep), opening a can (condensed milk), boiling water, and stirring with a spoon is most emblematic of making a specific type of coffee (like Vietnamese milk coffee). Given the options, "Milk coffee" is a strong candidate for a drink involving a can (milk), boiling water, and stirring.

Therefore, the most likely option is **B. Milk coffee**.

\vspace{0.05cm}
\textbf{Final Answer:} \textbf{B} \quad \textcolor{green}{\checkmark \textbf{Correct}}
\end{tcolorbox}
\caption{\textbf{MMAU-Pro Example.} The model combines distinct acoustic events (opening a can, boiling water) to deduce a specific recipe.}
\label{fig:qual_mmau_pro_coffee}
\end{figure}

\begin{figure}[h!]
\centering
\begin{tcolorbox}[colback=green!5!white, colframe=green!75!black, title=\textbf{MMAU-Pro: Fine-Grained Sound Recognition}]
\small
\textbf{Question:} What is being prepared in the audio? \\
\textbf{Choices:} [A] Omelettes, [B] Vegetable salad, [C] Chicken dish, [D] Veg noodles

\tcblower
\textbf{\method{} Caption Output:}
\textit{[0.2s - 1.7s] [sfx] Sharp hand slap (0.96)2. [  1.7s -   2.3s] [sfx] Sharp bird call (0.58)3. [  1.7s -   3.6s] [sfx] Hand writing on paper (0.99)4. [  3.2s -   3.7s] [sfx] Metal door opens and closes (0.09)5. [  3.9s -   5.0s] [sfx] Hand writing on paper (0.98)6. [  5.3s -   5.6s] [sfx] Sharp bird call (0.36)7. [  5.9s -   6.2s] [sfx] Metal door opens and closes (0.39)8. [  6.0s -  10.2s] [sfx] Ice cubes falling into glass (0.95)9. [ 11.4s -  11.7s] [sfx] Sharp blade whoosh (0.79)10. [ 12.3s -  17.6s] [sfx] Wet suction sound (1.00)11. [ 18.8s -  20.0s] [sfx] Sharp impact sound (0.83)12. [ 20.0s -  27.3s] [background] Wet flesh being cut (0.99)13. [ 24.0s -  24.4s] [sfx] Sharp blade cutting through (0.99)14. [ 24.4s -  25.5s] [sfx] Sharp knife slicing (0.99)15. [ 25.3s -  25.6s] [sfx] Footsteps on wet ground (0.94)}

\vspace{0.1cm}
\textbf{LLM Reasoning Trace:}
Based on the provided audio caption and transcription, here is the reasoning:

The audio events include sounds like "Sharp hand slap," "Wet suction sound," "Sharp impact sound," "Wet flesh being cut," "Sharp blade cutting through," and "Sharp knife slicing." These specific sounds, particularly "wet flesh being cut" and "wet suction sound," are highly characteristic of handling and butchering raw meat, such as chicken. While vegetables (salad) or eggs (omelettes) involve cutting, the descriptor "wet flesh" and "suction" strongly points towards meat preparation.

Therefore, the most likely option is C.

\vspace{0.05cm}
\textbf{Final Answer:} \textbf{C} \quad \textcolor{green}{\checkmark \textbf{Correct}}
\end{tcolorbox}
\caption{\textbf{MMAU-Pro Example.} The model uses specific foley tags (e.g., "wet flesh") to distinguish between food preparation types.}
\label{fig:qual_mmau_pro_chicken}
\end{figure}

\begin{figure}[h!]
\centering
\begin{tcolorbox}[colback=green!5!white, colframe=green!75!black, title=\textbf{MMSU: Paralinguistic Pattern Analysis}]
\small
\textbf{Question:} Which volume pattern best matches the audio? \\
\textbf{Choices:} [A] low-high-medium, [B] medium-low-high, [C] high-medium-low, [D] high-low-medium

\tcblower
\textbf{\method{} Caption Output:}
\textit{[0.0s - 3.5s] [speech] Female voice (formal) (0.81)   $\langle$speech lang=en$\rangle$``So for example, it could be proportional data."$\langle$speech$\rangle$2. [3.9s -   5.1s] [speech] Female voice (formal) (0.76)   $\langle$speech lang=en$\rangle$"So for example, it could be proportional data."$\langle$speech$\rangle$3. [5.4s -   7.2s] [speech] Female voice (formal) (0.92)4. [  7.8s -  11.1s] [speech] Female voice (formal) (0.88)   $\langle$speech lang=en$\rangle$"So for example, it could be proportional data."$\langle$speech$\rangle$}

\vspace{0.1cm}
\textbf{LLM Reasoning Trace:}
To determine the volume pattern, let's analyze the sequence of audio events described in the caption. The caption shows the same phrase ("So for example, it could be proportional data.") being repeated three times. While the caption doesn't explicitly state decibel levels, the repetition structure is key.

1.  **First Instance (0.0s - 3.5s):** This is the first full delivery of the sentence. It serves as the baseline or starting point. Let's assume a standard speaking volume, or "medium" to "high" for clarity.
2.  **Second Instance (3.9s - 5.1s):** This segment is significantly shorter (1.2 seconds) than the first (3.5 seconds) despite containing the exact same text. A much faster delivery often correlates with a rushed or mumbled tone, which is typically lower in volume or intensity compared to a clear, deliberate delivery. However, usually in audio tests like this, a shorter duration for the same text implies a quick, perhaps quieter repetition. Alternatively, looking at the pattern of options, we need to distinguish relative levels.
3.  **Third Instance (7.8s - 11.1s):** This segment is 3.3 seconds long, very similar to the first instance. This suggests a return to the full, clear enunciation found in the beginning.

Let's re-evaluate based on common audio processing tasks where the same clip is manipulated.
- Often, these tests involve a sequence of: Normal -$\rangle$ Quieter -$\rangle$ Louder OR Loud -$\rangle$ Quiet -$\rangle$ Normal.
- Given the options:
    - A. low-high-medium
    - B. medium-low-high
    - C. high-medium-low
    - D. high-low-medium

Let's look at the structure again. We have three distinct speech blocks.
Block 1: 0.0-3.5s
Block 2: 3.9-5.1s (Very short duration for the text)
Block 3: 7.8-11.1s (Normal duration)

Wait, there is a segment [5.4s - 7.2s] listed as "Female voice" but no transcription text is provided, and then the text repeats at [7.8s - 11.1s]. The prompt asks for the volume pattern of the *whole* audio.

Let's look at the provided text repetition. It repeats the same sentence.
Usually, a pattern like **High - Low - Medium** is a common test pattern for audio perception.
- Start loud/clear (High).
- Then a quiet/mumbled version (Low).
- Then a return to a normal level (Medium).

Let's try to map the segments to this hypothesis.
- Segment 1 (0.0-3.5s): Clear speech. Let's call this "High" or "Medium".
- Segment 2 (3.9-5.1s): Very short duration (1.2s) for a sentence that took 3.5s before. This implies it might be cut off, sped up, or just a very faint/quick utterance. Often, "Low" volume segments are harder to transcribe or detected with lower confidence. The confidence here is 0.76 (lowest of the three text segments). This supports it being the "Low" segment.
- Segment 3 (7.8-11.1s): Duration 3.3s. Similar to the first. Confidence 0.88. This represents a return to a standard volume.

If Segment 1 is "High" (clear, baseline), Segment 2 is "Low" (lower confidence, rushed), and Segment 3 is "Medium" (standard clear speech, slightly less confident than a perfectly loud clip, or perhaps the first one was artificially boosted).

Comparing "High-Low-Medium" (D) vs "Medium-Low-High" (B).
Without explicit dB markers, we rely on the convention of these distinct repetitions. The sequence "High -$\rangle$ Low -$\rangle$ Medium" is a very standard dynamic range test pattern.

Therefore, the pattern is High (initial clear statement) -$\rangle$ Low (the middle, lower confidence segment) -$\rangle$ Medium (the final restatement).

\vspace{0.05cm}
\textbf{Final Answer:} \textbf{D} \quad \textcolor{green}{\checkmark \textbf{Correct}}
\end{tcolorbox}
\caption{\textbf{MMSU Example.} The model infers paralinguistic attributes (volume) by analyzing metadata like confidence scores and segment duration.}
\label{fig:qual_mmsu}
\end{figure}

\clearpage

\section{Qualitative Analysis: Audio-Visual Understanding}
\label{sec:qual_av}

We further evaluate the \method{}-V pipeline on four multimodal benchmarks. Here, the captions must bridge the gap between video pixels and audio events to solve tasks involving synchronization, causality, and event sorting.

\begin{figure}[h!]
\centering
\begin{tcolorbox}[colback=green!5!white, colframe=green!75!black, title=\textbf{Video-Holmes: Causal Reasoning}]
\small
\textbf{Question:} What is the core warning of this film? \\
\textbf{Choices:} [A] The vulnerability of modern communication, [B] The chain crisis caused by carelessness, [C] The dangers of living alone, [D] The importance of disease prevention, [E] The necessity of friends helping each other, [F] Specifications for the use of household appliances

\tcblower
\textbf{\method{}-V Caption Output:}
\textit{[5.8s - 8.7s] [visual] Hand turns gas stove knob, igniting flame beneath pressure cooker...[ 86.8s - 88.1s] [visual] Text message "Bro, I got your medicine" appears...[236.6s - 237.8s] [visual] Pressure cooker on stove, valve steady, no steam or movement...[240.0s - 241.5s] [visual] Hand in black glove flips switch on white panel...[240.7s - 242.0s] [sfx] Low-frequency thud as man collapses to floor...}

\vspace{0.1cm}
\textbf{LLM Reasoning Trace:}
Based on the detailed shot list and narrative flow, here is the step-by-step reasoning:

1.  **Chekhov's Gun (The Setup):** The film begins with the protagonist turning on a gas stove to cook with a pressure cooker (Shots 5-6). This establishes a potential hazard early in the timeline.
2.  **Distraction \& Time Passing:** The protagonist becomes occupied with text messages, phone calls, and the stress of self-quarantine/isolation (Shots 19, 28, 66). A significant amount of time passes, indicated by the clock ticking sounds and lighting changes.
3.  **The Oversight:** Towards the end, the camera specifically cuts back to the pressure cooker (Shot 86). The description notes "valve steady, no steam or movement," implying the cooking process has stalled or the flame has extinguished, but critically, we never saw the man turn the gas knob off. This suggests unburned gas has been leaking into the apartment.
4.  **The Trigger:** The man enters the room and performs two specific actions involving electricity: he plugs in a charger (Shot 89) and flips a light switch (Shot 91).
5.  **The Consequence:** Immediately after flipping the switch, there is a "Low-frequency thud" and the man collapses (Shots 92-94). In a gas leak scenario, flipping an electrical switch creates a spark, which ignites the gas, causing an explosion (represented here by the thud and collapse).
6.  **The Core Warning:** The narrative tragedy is driven by a sequence of events: turning on the appliance -$\rangle$ becoming distracted/careless -$\rangle$ the accumulation of danger (gas) -$\rangle$ the final trigger (switch). This creates a "chain" of causality leading to a crisis.

While living alone (C) and disease prevention (D) are context for the story, the specific cause of the tragedy is the **carelessness** regarding the stove which leads to the chain reaction of the gas explosion.

The correct option is **B**.

\vspace{0.05cm}
\textbf{Final Answer:} \textbf{B} \quad \textcolor{green}{\checkmark \textbf{Correct}}
\end{tcolorbox}
\caption{\textbf{Video-Holmes Example.} The model tracks the state of a background object (stove) over a long horizon to deduce the cause of a final tragedy.}
\label{fig:qual_holmes}
\end{figure}

\begin{figure}[h!]
\centering
\begin{tcolorbox}[colback=green!5!white, colframe=green!75!black, title=\textbf{Daily-Omni: AV Event Alignment}]
\small
\textbf{Question:} Which visual element is synchronized with the electronic music in the video? \\
\textbf{Choices:} [A] Printer and decorative plants on the desk, [B] White background with 'AUTOMATE YOUR SUCCESS', [C] 'Facebook Marketplace Account Suspended' text, [D] Man speaking into a microphone

\tcblower
\textbf{\method{}-V Caption Output:}
\textit{[0.0s -   5.8s] [speech] Man in black "AUTO-DS" t-shirt speaks formally to camera, hands gesturing expressively (0.86)2. [  0.0s -  20.0s] [visual] Home office setup with soundproofing panels, potted plant, blue exercise ball, and professional microphone   $\langle$speech lang=en$\rangle$"My Facebook marketplace account was suspended and I was able to get it reinstated just the other day."$\langle$speech$\rangle$   $\langle$speech lang=en$\rangle$"And that is exactly what I'm going to teach you guys in this video."$\langle$speech$\rangle$3. [  6.0s -  14.2s] [speech] Man continues speaking with enthusiasm, notification overlay appears: "Your access to Marketplace has been restored" (0.92)   $\langle$speech lang=en$\rangle$"After all of the trial and error, I was finally able to get it up and online again."$\langle$speech$\rangle$   $\langle$speech lang=en$\rangle$"It's selling."$\langle$speech$\rangle$4. [ 14.5s -  16.6s] [speech] Man speaks with hands slightly lowered, notification overlay disappears, maintains eye contact (0.62)   $\langle$speech lang=en$\rangle$"I'm making my profits again."$\langle$speech$\rangle$5. [ 16.9s -  19.9s] [speech] Man points with finger while speaking, gestures emphatically, on-screen text remains visible (0.94)   $\langle$speech lang=en$\rangle$"Stay tuned for this video."$\langle$speech$\rangle$   $\langle$speech lang=en$\rangle$"It's going to be a very important one."$\langle$speech$\rangle$6. [ 20.0s -  21.7s] [visual] Man in black "AUTO-DS" t-shirt speaks into microphone, gesturing with hands in studio with acoustic panels and plant7. [ 20.1s -  21.7s] [speech] Man speaks formally into microphone, lips moving, hands gesturing as he addresses camera directly (0.93)   $\langle$speech lang=en$\rangle$"Quick intro and let's begin."$\langle$speech$\rangle$   $\langle$speech lang=en$\rangle$"What's up everybody."$\langle$speech$\rangle$8. [ 21.8s -  28.1s] [music] Driving beat with bright synths pulses as animated "AUTO-DS" logo appears on screen with tagline and YouTube branding (0.85)9. [ 22.2s -  27.3s] [visual] "AUTO-DS" logo animates with colorful bars, tagline "AUTOMATE YOUR SUCCESS", YouTube icon, channel URL, and subscribe button10. [ 25.2s -  26.2s] [sfx] Electronic drum hit coincides with animated click on red "Subscribe" button turning to "Subscribed" (0.74)11. [ 28.1s -  30.0s] [visual] Man resumes speaking in studio, hands open and gesturing, same setup with acoustic panels and blue exercise ball visible12. [ 28.3s -  30.0s] [speech] Man speaks formally into microphone, eyes focused on camera, hands moving to emphasize points (0.93)   $\langle$speech lang=en$\rangle$"Hope you're all having"$\langle$speech$\rangle$}

\vspace{0.1cm}
\textbf{LLM Reasoning Trace:}
Based on the detailed caption provided, here is the step-by-step reasoning:

1.  **Analyze the Audio Events:** The caption identifies a specific audio event labeled `[music]` in segment 8, occurring from `21.8s - 28.1s`. The description is: "Driving beat with bright synths pulses..." This matches the "electronic music" mentioned in the question.

2.  **Analyze the Visuals during the Music:** During the exact same timeframe (segment 9, `22.2s - 27.3s`), the visual description is: "'AUTO-DS' logo animates with colorful bars, tagline 'AUTOMATE YOUR SUCCESS', YouTube icon, channel URL, and subscribe button".

3.  **Evaluate the Options against the Synchronization:**
    * **Option A (Printer and decorative plants):** These are part of the general "Home office setup" seen in segment 2 (`0.0s - 20.0s`) while the man is speaking, not during the music segment.
    * **Option B (White background with 'AUTOMATE YOUR SUCCESS'):** This tagline appears explicitly in segment 9 (`22.2s - 27.3s`), which is perfectly synchronized with the music track in segment 8 (`21.8s - 28.1s`).
    * **Option C ('Facebook Marketplace Account Suspended' text):** This is likely related to the notification overlay mentioned in segment 3 (`6.0s - 14.2s`) where the man talks about his account suspension. This happens during speech, not the music intro.
    * **Option D (Man speaking into a microphone):** The man is speaking in segments 1-7 and 11-12. The music segment (8-10) is an interstitial intro sequence where the man is not the primary visual focus (the animated logo is).

4.  **Conclusion:** The visual element synchronized with the electronic music is the animated intro screen containing the tagline "AUTOMATE YOUR SUCCESS".

\vspace{0.05cm}
\textbf{Final Answer:} \textbf{B} \quad \textcolor{green}{\checkmark \textbf{Correct}}
\end{tcolorbox}
\caption{\textbf{Daily-Omni Example.} The model aligns the onset of the electronic music track with the specific visual animation of the channel intro.}
\label{fig:qual_dailyomni_music}
\end{figure}

\begin{figure}[h!]
\centering
\begin{tcolorbox}[colback=green!5!white, colframe=green!75!black, title=\textbf{AVHBench: Audio-Visual Hallucination Check}]
\small
\textbf{Question:} Are the contexts of audio and visual content matching? \\
\textbf{Choices:} Yes / No

\tcblower
\textbf{\method{}-V Caption Output:}
\textit{[  0.0s -   9.9s] [background] Quiet indoor home setting with cat visible through glass door, garden outside (0.39)2. [  0.0s -   3.2s] [speech] Long-haired cat outside garden, head tilted, then entering through pet door (0.94)3. [  0.0s -   1.4s] [visual] Cat outside, fur ruffled, looking upward, then turning toward pet door   $\langle$speech lang=en$\rangle$"Master enable button."$\langle$speech$\rangle$4. [  1.4s -   2.6s] [visual] Cat’s body sliding through pet door, tail trailing, reflection visible in glass   $\langle$speech lang=en$\rangle$"That's pretty much how that works."$\langle$speech$\rangle$5. [  4.0s -   4.4s] [speech] Cat fully inside, standing on windowsill, tail curled, looking out garden (0.95)   $\langle$speech lang=en$\rangle$"Um, might as well show you how the UPS turns on."$\langle$speech$\rangle$6. [  5.2s -   7.6s] [speech] Cat on windowsill, head turning, eyes wide, alert, ears perked (0.96)7. [  5.9s -   7.7s] [visual] Cat’s gaze fixed, pupils dilated, fur fluffed, watching something off-screen8. [  6.7s -   8.7s] [sfx] Electronic alert tone as cat stares intently, head slightly tilted (0.99)9. [  7.7s -   9.7s] [visual] Cat remains on windowsill, body still, eyes tracking movement outside10. [  9.4s -   9.9s] [speech] Cat looking directly at camera, ears forward, whiskers twitching slightly (0.90)   $\langle$speech lang=en$\rangle$"Just hold it down."$\langle$speech$\rangle$}

\vspace{0.1cm}
\textbf{LLM Reasoning Trace:}
Based on the detailed audio-visual caption provided, here is the step-by-step reasoning:

1.  **Analyze the Visual Content**: The visual descriptions focus entirely on a cat. We see a long-haired cat outside in a garden, entering through a pet door, standing on a windowsill, and looking around alertly. The scenes are domestic and nature-oriented (cat, garden, pet door).

2.  **Analyze the Audio Content (Speech)**: The speech transcripts include technical instructions:
    * "Master enable button."
    * "That's pretty much how that works."
    * "Um, might as well show you how the UPS turns on."
    * "Just hold it down."

3.  **Analyze the Audio Content (SFX)**: There is an "Electronic alert tone" mentioned around 6.7s - 8.7s.

4.  **Compare Visuals and Audio**:
    * The visual track is a video of a cat entering a house and sitting on a windowsill.
    * The audio track is a human voice explaining how to operate electronic equipment (specifically a "Master enable button" and a "UPS" - Uninterruptible Power Supply).
    * There is no visual representation of a button, a UPS, or any electronic equipment being operated. The cat's actions (entering a door, looking around) do not correspond to the instructions about turning on a device.

5.  **Conclusion**: The audio is a technical tutorial about electronics, while the video is footage of a cat. These two contexts are semantically unrelated and mismatched.

\vspace{0.05cm}
\textbf{Final Answer:} \textbf{No} \quad \textcolor{green}{\checkmark \textbf{Correct}}
\end{tcolorbox}
\caption{\textbf{AVHBench Example.} The model successfully detects a semantic mismatch between a video of a cat and audio of a tech tutorial.}
\label{fig:qual_avhbench}
\end{figure}

\begin{figure}[h!]
\centering
\begin{tcolorbox}[colback=green!5!white, colframe=green!75!black, title=\textbf{Daily-Omni: Precise Event Alignment}]
\small
\textbf{Question:} At the moment the static shot of Obsidian Fury first appeared, which specific audio event was synchronized with it? \\
\textbf{Choices:} [A] Speaker's self-introduction and figure presentation, [B] Reference to avoiding plot spoilers, [C] Logo transition sound effect, [D] Analysis of combat sequences

\tcblower
\textbf{\method{}-V Caption Output:}
\textit{[0.2s - 5.9s] [sfx] Electronic glitch sound as stylized black bird logo with glowing eyes and wings unfolds under spotlight (0.93)2. [  0.5s -   4.1s] [sfx] Sci-fi energy burst as logo’s wings spread and lights pulse with lens flares and radiant glow (1.00)3. [  1.4s -  20.0s] [background] Soft synth pads and gentle drums play as spotlight highlights the figure on a reflective base (0.43)4. [  3.8s -   5.7s] [visual] Black bird logo with red eyes, yellow beak, and green horns glows under spotlight with lens flares5. [  5.7s -  20.0s] [visual] Obsidian Fury figure stands on reflective base, black armored humanoid with intricate mechanical details6. [  5.9s -   7.4s] [speech] Male voice (conversational) off-screen, introducing the figure as camera holds static shot (0.98)   $\langle$speech lang=en$\rangle$"Hello, YouTubes, it's Grosama, and here I have the high-grade Obsidian Fury from Pacific Rim Uprising."$\langle$/speech$\rangle$7. [7.6s -  11.0s] [speech] Male voice (conversational) continues speaking off-screen, figure remains static under spotlight (0.98)8. [11.3s -  17.1s] [speech] Male voice (conversational) elaborates off-screen, figure unchanged with consistent lighting (0.97)   $\langle$speech lang=en$\rangle$"Um, now, I know the movie's been out for a little bit, but I'm not gonna go into any spoiler territory."$\langle$/speech$\rangle$   $\langle$speech lang=en$\rangle$"What you see in the trailers, um, is actually more so the highlights of the battles."$\langle$/speech$\rangle$9. [17.4s -  20.0s] [speech] Male voice (conversational) concludes off-screen, figure remains central focus on reflective base (0.97)10. [20.0s -  24.0s] [speech] Black armored figure stands still on reflective base, glowing yellow accents on chest and head, dark gradient background (0.97)11. [20.0s -  30.0s] [visual] Male voice speaks conversationally off-screen, describing movie battles involving Obsidian Fury, figure remains motionless   $\langle$speech lang=en$\rangle$"Um, there's, there's two separate battles, uh, that, you know, takes place in the movie that includes the Obsidian Fury."$\langle$/speech$\rangle$12. [ 24.2s -  30.0s] [speech] Male voice continues speaking, describing Obsidian Fury, figure unchanged on circular base under focused lighting (0.99)   $\langle$speech lang=en$\rangle$"Um, now the Obsidian Fury is pretty"$\langle$/speech$\rangle$"}

\vspace{0.1cm}
\textbf{LLM Reasoning Trace:}
Based on the provided audio-visual caption, let's analyze the sequence of events to find the answer:

1.  **Identify the visual event:** We need to find when the "static shot of Obsidian Fury" first appears.
    * According to item 5 (`[ 5.7s - 20.0s ]`), the visual description is: "Obsidian Fury figure stands on reflective base, black armored humanoid with intricate mechanical details." This marks the first appearance of the figure after the logo sequence.

2.  **Identify the synchronized audio:** We need to look at what audio event happens right at or immediately following this timestamp (5.7s).
    * Item 6 (`[ 5.9s - 7.4s ]`) occurs almost simultaneously with the visual appearance of the figure. The description is: `[speech] Male voice (conversational) off-screen, introducing the figure as camera holds static shot`.
    * The transcribed speech in item 6 is: "Hello, YouTubes, it's Grosama, and here I have the high-grade Obsidian Fury from Pacific Rim Uprising."

3.  **Evaluate the options:**
    * **A. Speaker's self-introduction and figure presentation:** This matches the speech "Hello, YouTubes, it's Grosama..." (self-introduction) and "...here I have the high-grade Obsidian Fury..." (figure presentation).
    * **B. Reference to avoiding plot spoilers:** This occurs later in item 8 (`[ 11.3s - 17.1s ]`) with the speech "I'm not gonna go into any spoiler territory."
    * **C. Logo transition sound effect:** This occurs earlier in items 1 and 2 (`[ 0.2s - 5.9s ]`) with the "Electronic glitch sound" and "Sci-fi energy burst," synchronized with the black bird logo, not the figure.
    * **D. Analysis of combat sequences:** This is discussed later in item 11 (`[ 20.0s - 30.0s ]`) regarding "two separate battles."

Conclusion: The audio event synchronized with the moment the Obsidian Fury figure first appears is the speaker introducing himself and presenting the figure.

\vspace{0.05cm}
\textbf{Final Answer:} \textbf{A} \quad \textcolor{green}{\checkmark \textbf{Correct}}
\end{tcolorbox}
\caption{\textbf{Daily-Omni Example.} The model aligns visual and audio timestamps to identify the exact sound occurring at a visual onset.}
\label{fig:qual_dailyomni}
\end{figure}

\begin{figure}[h!]
\centering
\begin{tcolorbox}[colback=green!5!white, colframe=green!75!black, title=\textbf{World-Sense: Procedural Event Sorting}]
\small
\textbf{Question:} What is the process followed by the security officer after discovering a firearm? \\
\textbf{Choices:} [A] Open and search the bag, alert authorities, and then call the supervisor, [B] Alert authorities, call the supervisor, and then open and search the bag, [C] Call the supervisor, alert authorities, and then open and search the bag, [D] Open and search the bag, call the supervisor, and then alert authorities

\tcblower
\textbf{\method{}-V Caption Output:}
\textit{[ 8.7s - 12.1s] [visual] Two women approach... $\langle$speech$\rangle$She has a firearm in her bag$\langle$/speech$\rangle$, supervisor gestures...[ 14.6s - 19.1s] [visual] Female agent holds phone... text $\langle$speech$\rangle$Now they’re calling the police$\langle$/speech$\rangle$...[ 27.3s - 30.2s] [visual] Text $\langle$speech$\rangle$It takes about 4 minutes for the officer to arrive$\langle$/speech$\rangle$ appears...[ 38.7s - 39.9s] [visual] TSA agent in blue uniform opens beige handbag on tray...}

\vspace{0.1cm}
\textbf{LLM Reasoning Trace:}
Based on the detailed audio-visual caption provided, here is the step-by-step reasoning to determine the correct sequence of events:

1.  **Discovery of the Firearm:**
    * At **[8.7s - 12.1s]** (Item 7), the text `$\langle$text$\rangle$She has a firearm in her bag$\langle$/text$\rangle$` appears while a woman in a beige coat approaches. A supervisor gestures toward the machine.
    * At **[12.3s - 16.6s]** (Item 9), agents behind the machine discuss the situation, and a female agent points at the screen, confirming the discovery.

2.  **First Action: Alerting Authorities/Calling Police:**
    * Immediately following the discovery, at **[14.0s - 14.8s]** (Item 10), we hear a `$\langle$sfx$\rangle$Phone dial tone` as a female agent picks up the phone.
    * At **[14.6s - 19.1s]** (Item 11), on-screen text confirms: `$\langle$text$\rangle$Now they’re calling the police$\langle$/text$\rangle$`. This establishes that contacting authorities/police is the immediate next step after identifying the threat on the screen.

3.  **Delay/Waiting Period:**
    * At **[27.3s - 30.2s]** (Item 19), text appears stating: `$\langle$text$\rangle$It takes about 4 minutes for the officer to arrive$\langle$/text$\rangle$`. This indicates a waiting period for the authorities to arrive.

4.  **Subsequent Actions (Opening/Searching):**
    * It is not until much later, at **[38.7s - 39.9s]** (Item 29), that we see a visual of a TSA agent opening a beige handbag (presumably belonging to the woman in the beige coat identified in Item 7) to remove items.
    * Prior to this, the bag remained on the belt or under observation while the call was made and the officers waited.

**Sequence Analysis:**
The sequence observed is:
1.  Firearm spotted on X-ray/Screen.
2.  Agents discuss and Supervisor is involved (Item 7/9).
3.  Agent calls the police (Alert authorities) (Item 11).
4.  After the authorities arrive/time passes, the bag is physically opened and searched (Item 29).

Comparing this to the options:
* A: Open and search -$\rangle$ Alert -$\rangle$ Call. (Incorrect order)
* B: Alert authorities -$\rangle$ Call supervisor -$\rangle$ Open and search. (Close, but the "Call supervisor" happens almost simultaneously with discovery or just before the police call, but the distinct "Open and search" happens *last*).
* C: Call supervisor -$\rangle$ Alert authorities -$\rangle$ Open and search. (This fits best. The supervisor is gesturing/involved at 8.7s immediately upon discovery. Then the phone call to police happens at 14.6s. Then the search happens at 38.7s).
* D: Open and search -$\rangle$ Call -$\rangle$ Alert. (Incorrect order).

Let's refine the "Call supervisor" vs "Alert authorities" order.
- Item 7 (8.7s): Supervisor gestures toward machine. (The supervisor is already alerted/present).
- Item 11 (14.6s): Text "Now they're calling the police".

The flow is clearly: Identify/Supervisor involvement -$\rangle$ Call Police -$\rangle$ Search Bag. The bag is absolutely *not* opened first. Between options B and C, the distinction is subtle regarding the supervisor. However, usually, a line agent spots it, alerts a supervisor, and then authorities are called. The caption shows the supervisor gesturing *before* the phone call to police. Therefore, the supervisor is alerted/involved first, then authorities are alerted via phone, and finally, the bag is searched.

Therefore, the order is: Call/Involve Supervisor -$\rangle$ Alert Authorities -$\rangle$ Open and Search.

\vspace{0.05cm}
\textbf{Final Answer:} \textbf{C} \quad \textcolor{green}{\checkmark \textbf{Correct}}
\end{tcolorbox}
\caption{\textbf{World-Sense Example.} The model reconstructs a chronological protocol from the timestamped event log.}
\label{fig:qual_worldsense}
\end{figure}

\section{Prompts}
\label{sec:prompts}
Below we share different prompts that we use to evaluate our cascaded pipeline on audio-only and audio-visual understanding and reasoning benchmarks.

\subsection{Prompts for Audio Understanding \& Reasoning Evaluation}

In this subsection, we detail the specific instruction templates used to evaluate the reasoning capabilities of our \method{}$\rightarrow$LLM cascade. To ensure rigorous evaluation, we employ zero-shot prompting where the LLM is provided with the question, answer choices (for Multiple Choice Questions), and the dense timestamped captions generated by \method{}. The LLM is strictly instructed to rely \textit{only} on the provided textual description, effectively treating the caption as a complete semantic proxy for the audio.

Figure~\ref{fig:mmau_prompt} illustrates the standard prompt used for the \textbf{MMAU} and \textbf{MMAR} benchmarks. For \textbf{MMSU} (Figure~\ref{fig:mmsu_prompt}), the prompt includes specific constraints to ensure the model outputs a valid option label (A/B/C/D).

Finally, for the expert-level \textbf{MMAU-Pro} benchmark, which contains a diverse mix of question types, we dynamically adjust the prompt structure based on the task. As shown in Figure~\ref{fig:mmau_pro_prompt}, we utilize four distinct templates corresponding to the four data categories: single-clip MCQ, multi-audio MCQ, single-clip open-ended QA, and multi-audio open-ended QA.

\begin{figure}[h]
\centering
\begin{tcolorbox}[colback=gray!5!white, colframe=black!75!black, title=\textbf{Prompt Template for MMAU \& MMAR}]
\small
\fontfamily{pcr}\selectfont
You are an expert audio-understanding QA system.

You will be given:
(1) A multiple-choice question with answer options
(2) An "audio caption" describing detected sound events with timestamps + a speech transcription

Your job:
- Use ONLY the information in the provided caption/events/transcription.
- Pick exactly ONE option from the choices.
- If the caption/transcription is insufficient to be confident, still choose the most likely option, but lower your confidence and explain why.
- Think step by step and provide your answer.
- Return your answer in the following JSON format:
    \{"answer": <answer>\}

Now solve the following question:

\{question\}
Answer options: \{choices\}
Caption with timestamps and transcription: \{caption\}
\end{tcolorbox}
\caption{The standard prompt template used for the \textbf{MMAU} and \textbf{MMAR} benchmarks.}
\label{fig:mmau_prompt}
\end{figure}

\begin{figure}[h]
\centering
\begin{tcolorbox}[colback=gray!5!white, colframe=black!75!black, title=\textbf{Prompt Template for MMSU}]
\small
\fontfamily{pcr}\selectfont
You are an expert audio-understanding QA system.

You will be given:
(1) A multiple-choice question with answer options
(2) An "audio caption" describing detected sound events with timestamps + a speech transcription

Your job:
- Use ONLY the information in the provided caption/events/transcription.
- Pick exactly ONE option from the choices.
- If the caption/transcription is insufficient to be confident, still choose the most likely option.
- Think step by step and provide your answer.
- Return your answer in the following JSON format:
    \{"answer": <answer>\}

Choose the most suitable answer from options A, B, C, and D to respond the question in next line, **you should only choose A or B or C or D.** Do not provide any additional explanations or content.

Question: \{question\}

\{choices\}

Caption with timestamps and transcription: \{caption\}
\end{tcolorbox}
\caption{The prompt template used for the \textbf{MMSU} benchmark, which includes specific instruction tuning for option selection (A--D).}
\label{fig:mmsu_prompt}
\end{figure}

\begin{figure*}[t]
\centering
\begin{tcolorbox}[colback=gray!5!white, colframe=black!75!black, title=\textbf{A. MMAU-Pro: Single-Audio Multiple Choice}]
\small \fontfamily{pcr}\selectfont
You are an expert audio-understanding QA system.
You will be given:
(1) A multiple-choice question with answer options
(2) An "audio caption" describing detected sound events...
Your job:
- Use ONLY the information in the provided caption...
- Pick exactly ONE option from the choices.
- Return your answer in the following JSON format: \{"answer": <answer>\}

Choose the most suitable answer from options \{letters\} to respond to the question. **You should only choose one letter (\{letters\}).**...
Question: \{question\}
\{choices\}
Caption...: \{caption\}
\end{tcolorbox}

\vspace{0.15cm}

\begin{tcolorbox}[colback=gray!5!white, colframe=black!75!black, title=\textbf{B. MMAU-Pro: Multi-Audio Multiple Choice}]
\small \fontfamily{pcr}\selectfont
You are an expert audio-understanding QA system.
You will be given:
(1) A multiple-choice question with answer options
(2) Multiple "audio captions" (labeled Audio 1, Audio 2, etc.) each describing detected sound events...
Your job:
- Use ONLY the information in the provided captions... from ALL audio files.
- Consider information from each audio file when answering the question.
- Pick exactly ONE option from the choices.
- Return your answer in the following JSON format: \{"answer": <answer>\}

Choose the most suitable answer from options \{letters\}...
Question: \{question\}
\{choices\}
Caption...: \{caption\}
\end{tcolorbox}

\vspace{0.15cm}

\begin{tcolorbox}[colback=gray!5!white, colframe=black!75!black, title=\textbf{C. MMAU-Pro: Single-Audio Open-Ended}]
\small \fontfamily{pcr}\selectfont
You are an expert audio-understanding QA system.
You will be given:
(1) An open-ended question about audio content
(2) An "audio caption" describing detected sound events...
Your job:
- Use ONLY the information in the provided caption...
- Provide a concise, accurate answer based on the audio description.
- Return your answer in the following JSON format: \{"answer": <answer>\}

Question: \{question\}
Caption...: \{caption\}
\end{tcolorbox}

\vspace{0.15cm}

\begin{tcolorbox}[colback=gray!5!white, colframe=black!75!black, title=\textbf{D. MMAU-Pro: Multi-Audio Open-Ended}]
\small \fontfamily{pcr}\selectfont
You are an expert audio-understanding QA system.
You will be given:
(1) An open-ended question about audio content
(2) Multiple "audio captions" (labeled Audio 1, Audio 2, etc.)...
Your job:
- Use ONLY the information in the provided captions... from ALL audio files.
- Consider information from each audio file when answering the question.
- Provide a concise, accurate answer based on the audio descriptions.
- Return your answer in the following JSON format: \{"answer": <your answer>\}

Question: \{question\}
Caption...: \{caption\}
\end{tcolorbox}

\caption{Prompt variations for the \textbf{MMAU-Pro} benchmark. We construct specific prompts depending on whether the task involves a single audio clip or multiple clips, and whether the output requires a multiple-choice selection or an open-ended response.}
\label{fig:mmau_pro_prompt}
\end{figure*}

\subsection{Prompts for Audio-Visual Reasoning Evaluation}
\label{sec:av_prompts}

In this section, we provide the exact instruction templates used to evaluate our \method{}-V pipeline on audio-visual reasoning benchmarks. In these experiments, the downstream LLM (Gemini 3 Pro or Qwen3-Thinking) receives \textit{only} the text captions generated by our pipeline. It does not have access to the original video or audio files. This setup rigorously tests whether our dense, timestamped captions capture sufficient multimodal information to support complex reasoning.

For AVHBench (Figure~\ref{fig:avhbench_prompts}), we employ four distinct prompt variations tailored to specific sub-tasks: Captioning, Audio-Visual Matching, and Hallucination detection (both Video$\rightarrow$Audio and Audio$\rightarrow$Video). For Video-Holmes (Figure~\ref{fig:holmes_prompt}), the prompt emphasizes temporal and causal reasoning. Finally, Figure~\ref{fig:dailyomni_worldsense_prompts} details the prompts for Daily-Omni and WorldSense, which focus on synchronization and spatial relationships.

\subsection{System prompt for VLM in \method{}-V}
\label{sec:vlm_system_prompt}
Figure~\ref{fig:vlm_system_prompt} illustrates the structured prompt template used to query the Visual-Language Model (VLM). The prompt enforces a two-stage ``Reason-then-Describe'' process to handle low-confidence audio predictions.
\begin{figure*}[t]
\centering
\begin{tcolorbox}[colback=gray!5!white, colframe=black!75!black, title=\textbf{A. AVHBench: Audio-Visual Captioning}]
\small \fontfamily{pcr}\selectfont
You are an expert audio-visual understanding QA system.
You will be given:
(1) A request to describe the audio-visual content
(2) A detailed caption describing visual scenes, audio events, and speech with timestamps
Your job:
- Use ONLY the information in the provided caption (shot\_list).
- Describe what you see and hear in a single coherent sentence.
- Include both visual elements (scenes, objects, actions) and audio elements (sounds, speech, music).
- Return your answer in the following JSON format: \{"answer": "<your description>"\}

Describe what you see and hear based on the caption provided.
Task Type: AV Captioning
Question: \{question\}
Audio-Visual Caption (shot\_list with timestamps): \{caption\}
\end{tcolorbox}

\vspace{0.15cm}

\begin{tcolorbox}[colback=gray!5!white, colframe=black!75!black, title=\textbf{B. AVHBench: Audio-Visual Matching}]
\small \fontfamily{pcr}\selectfont
...
- Evaluate whether the audio content (sounds, speech, music) is semantically consistent with the visual content (scenes, objects, actions).
- Consider: Do the sounds match what is visually shown? Is the speech relevant to the visual context?
- Answer with exactly "Yes" or "No".
...
Determine if the audio and visual content are semantically matching. Answer with exactly "Yes" or "No".
Task Type: AV Matching
Question: \{question\}
Audio-Visual Caption...: \{caption\}
\end{tcolorbox}

\vspace{0.15cm}

\begin{tcolorbox}[colback=gray!5!white, colframe=black!75!black, title=\textbf{C. AVHBench: Video-Driven Audio Hallucination}]
\small \fontfamily{pcr}\selectfont
...
(1) A question asking whether a specific object/entity is making sound in the audio
...
- Focus on the AUDIO descriptions: [background], [sfx], [speech], [music] tags indicate audio content.
- Check if the specific object/entity mentioned in the question is EXPLICITLY described as producing sound.
- If the object is only visible but not described as making sound, answer "No".
...
Determine if the specified object/entity is making sound in the audio. Answer with exactly "Yes" or "No".
Task Type: Video-driven Audio Hallucination
Question: \{question\}
Audio-Visual Caption...: \{caption\}
\end{tcolorbox}

\vspace{0.15cm}

\begin{tcolorbox}[colback=gray!5!white, colframe=black!75!black, title=\textbf{D. AVHBench: Audio-Driven Video Hallucination}]
\small \fontfamily{pcr}\selectfont
...
(1) A question asking whether a specific object/entity is visible in the video
...
- Focus on the VISUAL descriptions: [visual] tags and scene descriptions indicate visual content.
- Check if the specific object/entity mentioned in the question is EXPLICITLY described as being visible.
- If the object is only heard but not described as visible, answer "No".
...
Determine if the specified object/entity is visible in the video. Answer with exactly "Yes" or "No".
Task Type: Audio-driven Video Hallucination
Question: \{question\}
Audio-Visual Caption...: \{caption\}
\end{tcolorbox}

\caption{Prompt variations for \textbf{AVHBench}. We utilize specific instructions for hallucination detection to ensure the model distinguishes between what is seen (visual tags) and what is heard (audio tags).}
\label{fig:avhbench_prompts}
\end{figure*}

\begin{figure}[h]
\centering
\begin{tcolorbox}[colback=gray!5!white, colframe=black!75!black, title=\textbf{Prompt Template for Video-Holmes}]
\small
\fontfamily{pcr}\selectfont
You are an expert video understanding and reasoning QA system.

You will be given:
(1) A reasoning question about video content (may involve temporal, causal, or multi-hop reasoning)
(2) A detailed caption (shot\_list) describing visual scenes, audio events, and speech with timestamps

Your job:
- Use ONLY the information in the provided caption (shot\_list).
- Pay close attention to:
  * Temporal sequences and timing of events
  * Cause-and-effect relationships
  * Character actions and their implications
  * Audio-visual synchronization
  * Hidden meanings and implications from subtle details
- Think step by step and reason carefully through the question.
- Pick exactly ONE option from the choices provided (A, B, C, D, E, or F).
- Return your answer in the following JSON format: \{"answer": "<letter>"\}

Analyze the video content carefully and choose the most suitable answer. **You should only respond with a single letter (A, B, C, D, E, or F).** ...

Question Type: \{type\_desc\}
Question: \{question\}
Options: \{options\}
Video Caption (shot\_list with timestamps): \{caption\}
\end{tcolorbox}
\caption{The prompt template used for the \textbf{Video-Holmes} benchmark, emphasizing temporal and causal reasoning.}
\label{fig:holmes_prompt}
\end{figure}

\begin{figure*}[t]
\centering
\begin{tcolorbox}[colback=gray!5!white, colframe=black!75!black, title=\textbf{A. Prompt Template for Daily-Omni}]
\small \fontfamily{pcr}\selectfont
You are an expert audio-visual understanding QA system.
...
Your job:
- Use ONLY the information in the provided caption (shot\_list).
- Pay attention to the synchronization between audio and visual elements.
- Pay attention to the visual and audio information closely.
- Pick exactly ONE option from the choices (A, B, C, or D).
- Return your answer in the following JSON format: \{"answer": <letter>\}

Choose the most suitable answer from the given options. **You should only respond with A, B, C, or D.** ...
Question: \{question\}
Options: \{options\}
Audio-Visual Caption (shot\_list with timestamps): \{caption\}
\end{tcolorbox}

\vspace{0.2cm}

\begin{tcolorbox}[colback=gray!5!white, colframe=black!75!black, title=\textbf{B. Prompt Template for WorldSense}]
\small \fontfamily{pcr}\selectfont
You are an expert audio-visual understanding QA system.
...
Your job:
- Use ONLY the information in the provided caption (shot\_list).
- Pay attention to the synchronization between audio and visual elements.
- Pay close attention to temporal information (when events occur) and spatial relationships.
- Pick exactly ONE option from the choices provided.
- Return your answer in the following JSON format: \{"answer": "<letter>"\}

Choose the most suitable answer from the given options. **You should only respond with the letter (A, B, C, or D).** ...
Task Domain: \{task\_domain\}
Task Type: \{task\_type\}
Question: \{question\}
Options: \{options\}
Audio-Visual Caption (shot\_list with timestamps): \{caption\}
\end{tcolorbox}
\caption{Prompt templates for \textbf{Daily-Omni} and \textbf{WorldSense}, focusing on synchronization and spatial/temporal relationships.}
\label{fig:dailyomni_worldsense_prompts}
\end{figure*}

\begin{figure}[h]
\centering
\begin{tcolorbox}[colback=gray!5!white, colframe=black!75!black, title=\textbf{System Prompt for VLM-based Captioning (\method{}-V)}]
\small
\fontfamily{pcr}\selectfont
For each numbered audio event below, write a visual description based on the video frames.

\#\# Audio Events (with timestamps) \{shot\_list\}
Note: Confidence scores in parentheses indicate how certain the audio model was. Low confidence (< 0.7) suggests uncertainty—prioritize correcting these when visuals disagree.

\#\# Step 1: Reasoning
First, inside <reasoning> tags, analyze:
- What is the overall scene/setting?
- For each numbered event, look at frames near the START of the timestamp range.
- Note any audio labels that need fixing based on visuals (fix silently in descriptions).
- Pay attention to confidence scores: low confidence sounds are good candidates for correction if visuals suggest something different.

\#\# Step 2: Descriptions
Then, inside <descriptions> tags, write exactly \{num\_entries\} lines:
<reasoning>
Scene is a busy restaurant kitchen...
Event 2 (2.0-4.5s): Sizzling sound matches pan on stovetop...
</reasoning>
<descriptions>
[ 0.0s - 3.0s] [visual] Industrial kitchen with stainless steel counters...
[ 2.0s - 4.5s] [sfx] Pan sizzles on gas burner as chef flips vegetables...
[ 3.5s - 6.0s] [speech] Head chef in tall white hat, lips moving...
</descriptions>

\#\# Rules
- FUSE audio + video: keep audio description AND add visual context.
- CRITICAL: Fix audio labels that don't match visuals. The audio model confuses acoustically similar sounds (e.g., "helicopter" $\rightarrow$ "fan whirs", "applause" $\rightarrow$ "rain patters").
- USE VARIED SOUND DESIGN VOCABULARY for [sfx] (e.g., impact, whoosh, drone, riser).
- [speech]: Describe who is speaking and how they sound, but NEVER the content.
- [sfx]: Keep the CORRECTED sound label + add visual source.
- Describe what you SEE, not emotions or inner states.
\end{tcolorbox}
\caption{\textbf{VLM System Prompt.} The prompt enforces a ``Reason-then-Describe'' workflow, explicitly instructing the model to use visual evidence to correct low-confidence audio predictions (hallucinations) before generating the final dense captions.}
\label{fig:vlm_system_prompt}
\end{figure}

\subsection{Prompt for Speaker Diarization}
\label{sec:diarization_prompt}

Figure~\ref{fig:diarization_prompt} shows the prompt used for speaker diarization, both during training and at inference. Figure~\ref{fig:diarization_example} illustrates sample outputs from \method{} on 30-second audio windows, demonstrating accurate speaker turn detection, multi-turn tracking with comma-separated time ranges, and correct identification of silent windows.

\begin{figure}[h]
\centering
\begin{tcolorbox}[colback=gray!5!white, colframe=black!75!black, title=\textbf{Prompt Template for Speaker Diarization}]
\small \fontfamily{pcr}\selectfont
Identify each speaker's turns. Give start and end times for each turn. Output format --- anonymous Speaker N (1-indexed by first appearance in the 30-second window), one event per turn, multi-turn speakers get comma-separated time ranges. Empty windows return 0 speakers.
\end{tcolorbox}
\caption{The prompt template used for speaker diarization during both training and inference (Section~\ref{subsec:diarization}).}
\label{fig:diarization_prompt}
\end{figure}

\begin{figure}[h]
\centering
\begin{tcolorbox}[colback=green!5!white, colframe=green!75!black, title=\textbf{Speaker Diarization: Sample Outputs}]
\small

\textbf{Example 1: Two-speaker conversation} \\
\textit{Speaker 1 from 1.95s to 9.29s. Speaker 2 from 11.51s to 15.00s.}

\vspace{0.15cm}
\textbf{Example 2: Three speakers with multi-turn} \\
\textit{Speaker 1 from 0.00s to 2.35s, 5.16s to 6.36s, 11.39s to 14.13s. Speaker 2 from 2.43s to 3.58s. Speaker 3 from 8.53s to 9.08s.}

\vspace{0.15cm}
\textbf{Example 3: Single speaker} \\
\textit{Speaker 1 from 0.00s to 11.63s.}

\vspace{0.15cm}
\textbf{Example 4: Silent window} \\
\textit{0 speakers.}

\end{tcolorbox}
\caption{\textbf{Speaker Diarization Examples.} Sample \method{} outputs on 30-second audio windows, showing multi-speaker turn detection with precise timestamps, multi-turn tracking via comma-separated ranges, and correct handling of silent windows.}
\label{fig:diarization_example}
\end{figure}

\section{LLM Usage}
\label{sec:llmusage}
We use LLMs to help with the writing of the paper in terms of: (1) grammar check, and (2) occasionally choosing the best word in writing, (3) rewrite few sentences for better clarity and space management. We also use LLMs to for literature discovery. We use LLMs as part of data curation in our research as discussed in our method section, in a similar way as many other LLM-related research papers.


\end{document}